\documentclass{article}
\usepackage[vmargin=1in,hmargin=1in]{geometry}
\usepackage{amsmath,amssymb,amsfonts,graphicx,amsthm,nicefrac,mathtools,bm}
\usepackage{hyperref}
\usepackage[numbers]{natbib}
\usepackage[english]{babel}
\usepackage{times}
\usepackage[T1]{fontenc}
\usepackage{bbm,mdframed}
\usepackage[dvipsnames]{xcolor}
\usepackage{xspace}
\usepackage{rotating,multirow}
\usepackage{graphics,tikz}

\usepackage{grffile}
\usepackage{subcaption}
\usepackage{epstopdf} 
\usepackage[ruled]{algorithm2e}
\usepackage{float}
\usepackage[toc,title,page]{appendix}
\newtheorem{theorem}{Theorem}

\newtheorem{lemma}{Lemma}

\theoremstyle{definition}

\theoremstyle{remark}

\makeatletter

\newtheorem*{rep@theorem}{\rep@title}
\newcommand{\newreptheorem}[2]
{\newenvironment{rep#1}[1]
	{\def\rep@title{#2 \ref{##1}} \begin{rep@theorem}}%
		{\end{rep@theorem}}}
\makeatother
\newreptheorem{theorem}{Theorem}
\newreptheorem{lemma}{Lemma}
\newreptheorem{corollary}{Corollary}
\newreptheorem{proposition}{Proposition}

\newcommand{\figref}[1]{Figure~\ref{fig:#1}}

\newcommand{\secref}[1]{Section~\ref{sec:#1}}

\newcommand{\lemref}[1]{Lemma~\ref{lem:#1}}

\newcommand{\thmref}[1]{Theorem~\ref{thm:#1}}

\newcommand{\eqnref}[1]{\eqnref{eq:#1}}

\newcommand{\PP}[1]{\textnormal{Pr}\!\left\{{#1}\right\}} 
\newcommand{\EE}[1]{\mathbb{E}\left[{#1}\right]} 

\newcommand{\EEst}[2]{\mathbb{E}\left[{#1}\ \middle| \ {#2}\right]} 
\newcommand{\PPst}[2]{\text{Pr}\!\left\{{#1}\ \middle| \ {#2}\right\}} 

\def\R{\mathbb{R}}

\def\C{\mathbb{C}}

\def\Ft{\mathcal{F}^{t-1}}
\def\Fj{\mathcal{F}^{j-1}}

\def\H0{\mathcal{H}^0}

\newcommand{\ident}{\mathbf{1}}

\newcommand{\ignore}[1]{}

\newcommand{\nulls}{\mathcal{H}_0}
\newcommand{\FDP}{\textnormal{FDP}}

\usepackage{fancyhdr}
\newcommand{\thedate}{\today}
\newcommand{\theauthor}{}
\newcommand{\thetitle}{ADDIS:
	adaptive discarding algorithms for online FDR control \\with conservative nulls}

\date{\thedate}
\author{\theauthor}
\title{\thetitle}

\fancypagestyle{plain}{%
	\fancyhf{}%
	\lhead{\thepage}
}

\usepackage[mmddyy]{datetime}

\newcommand{\fdp}{\textnormal{FDP}}
\newcommand{\fdr}{\textnormal{FDR}}
\newcommand{\mfdr}{\textnormal{mFDR}}
\newcommand{\fdphat}{\widehat{\textnormal{FDP}}}

\def\S{\mathcal{S}}
\def\C{\mathcal{C}}
\def\R{\mathcal{R}}
\def\F{\mathcal{F}}

\def\H{\mathcal{H}}
\def\N{\mathbb{N}}

\makeatletter
\newcommand{\dotfrac}[2]{
	\mathchoice
	{\ooalign{$\genfrac{}{}{0pt}{0}{#1}{#2}$\cr\leavevmode\cleaders\hb@xt@ .22em{\hss $\displaystyle\cdot$\hss}\hfill\kern\z@\cr}}
	{\ooalign{$\genfrac{}{}{0pt}{1}{#1}{#2}$\cr\leavevmode\cleaders\hb@xt@ .22em{\hss $\textstyle\cdot$\hss}\hfill\kern\z@\cr}}
	{\ooalign{$\genfrac{}{}{0pt}{2}{#1}{#2}$\cr\leavevmode\cleaders\hb@xt@ .22em{\hss $\scriptstyle\cdot$\hss}\hfill\kern\z@\cr}}
	{\ooalign{$\genfrac{}{}{0pt}{3}{#1}{#2}$\cr\leavevmode\cleaders\hb@xt@ .22em{\hss $\scriptscriptstyle\cdot$\hss}\hfill\kern\z@\cr}}
}
\makeatother

\newcommand{\defn}{\ensuremath{:\, =}}

\tikzstyle{none} = [rectangle, rounded corners, minimum width=0.5cm, minimum height=0.5cm,text centered, draw=black, fill=blue!20]
\tikzstyle{noneoff} = [rectangle, rounded corners, minimum width=0.5cm, minimum height=0.5cm,text centered, draw=black, fill=blue!5]

\tikzstyle{ada} = [rectangle, rounded corners, minimum width=0.5cm, minimum height=0.5cm, text centered, draw=black, fill=orange!20]
\tikzstyle{adaoff} = [rectangle, rounded corners, minimum width=0.5cm, minimum height=0.5cm, text centered, draw=black, fill=orange!5]

\tikzstyle{adadis} = [rectangle, rounded corners, minimum width=0.5cm, minimum height=0.5cm, text centered, draw=black, fill={rgb:red,10;green,20;yellow,54}]
\tikzstyle{adadisoff} = [rectangle, rounded corners, minimum width=0.5cm, minimum height=0.5cm, text centered, draw=black, fill=green!5]

\tikzstyle{arrow} = [thick,->,>=stealth, text width=3cm]
\tikzstyle{dotarrow} = [dashed, text width=3cm]

\tikzstyle{title} = [rectangle, rounded corners, minimum width=1cm, minimum height=1cm, text centered, draw=white, fill=none]

\begin{document}
	\author{
		Jinjin Tian, Aaditya Ramdas\\
		Department of Statistics and Data Science\\
		Carnegie Mellon University\\
		\texttt{jinjint@andrew.cmu.edu, aramdas@stat.cmu.edu}
	}
	\maketitle
	
	\begin{abstract}
Major internet companies routinely perform tens of thousands of A/B tests each year. Such large-scale sequential experimentation has resulted in a recent spurt of new algorithms that can provably control the false discovery rate (FDR) in a fully online fashion. However, current state-of-the-art adaptive algorithms can suffer from a significant loss in power if null $p$-values are conservative (stochastically larger than the uniform distribution), a situation that occurs frequently in practice. 
In this work, we introduce a new adaptive discarding method called ADDIS that provably controls the FDR and achieves the best of both worlds: it enjoys appreciable power increase over all existing methods if  nulls are conservative (the practical case), and rarely loses power if nulls are exactly uniformly distributed (the ideal case). 
We provide several practical insights on robust choices of tuning parameters, and extend the idea to asynchronous and offline settings as well. 
\end{abstract}

\section{Introduction}\label{sec:intro}

Rapid data collection is making the online testing of hypotheses increasingly essential, where a stream of hypotheses $H_1,H_2,\dots$ is tested sequentially one by one. On observing the data for the $t$-th test which is usually summarized as a $p$-value $P_t$, and  without knowing the outcomes of the future tests, we must make the decision of whether to reject the corresponding null hypothesis $H_t$ (thus proclaiming a ``discovery''). Typically, a decision takes the form $I(P_t \leq \alpha_t)$ for some $\alpha_t \in (0,1)$, meaning that  we reject the null hypothesis when the $p$-value is smaller than some threshold $\alpha_t$. An incorrectly rejected null hypothesis is called a false discovery. Let $\R(T)$ represent the set of rejected null hypotheses until time $T$, and $\nulls$ be the unknown set of true null hypotheses; then, $\R(T) \cap \nulls$ is the set of false discoveries. Then some natural error metrics are the false discovery rate (FDR), modified FDR (mFDR) and power, which are defined as
\begin{equation}\label{fdr}
  \fdr(T) \equiv \EE{\frac{|\nulls\cap\R(T)|}{|\R(T)|\vee 1}},\ \ \  \mfdr(T) \equiv \frac{ \EE{|\nulls\cap\R(T)|}} {\EE{|\R(T)|\vee 1}},\ \ \  \textnormal{power} \equiv \EE{\frac{|\nulls^{c}\cap \R(T)|}{|\nulls^{c}|}}.
\end{equation}
The typical aim is to maximize power, while have $\fdr(T) \leq \alpha$ at any time $T \in \N$, for some prespecified constant $\alpha\in(0,1)$.
It is well known that setting every $\alpha_t\equiv\alpha$ does not provide any control of the FDR in general. Indeed, the FDR can be as large as one in this case, see \cite[Section 1]{ramdas2017on} for an  example. This motivates the need for special methods for online FDR control (that is, for determining $\alpha_t$ in an online manner).

\paragraph{Past work.} \citet{foster2008alpha} proposed the first ``alpha-investing'' (AI) algorithm for online FDR control, which was later extended to the generalized alpha-investing methods (GAI) by \citet{aharoni2014generalized}. A particularly powerful GAI algorithm called LORD was proposed by \citet{javanmard2017on}. Soon after, \citet{ramdas2017on} proposed a modification called LORD++ that uniformly improved the power of LORD. 
Most recently, \citet{ramdas2018saffron} developed the ``adaptive'' SAFFRON algorithm, and alpha-investing is shown to be a special case of the more general SAFFRON framework. SAFFRON arguably represents the state-of-the-art, achieving significant power gains over all other algorithms including LORD++ in a range of experiments.

\begin{figure}[H]
	\centering
	\begin{subfigure}{0.4\textwidth}
		\centering
		\includegraphics[width=0.9\linewidth]{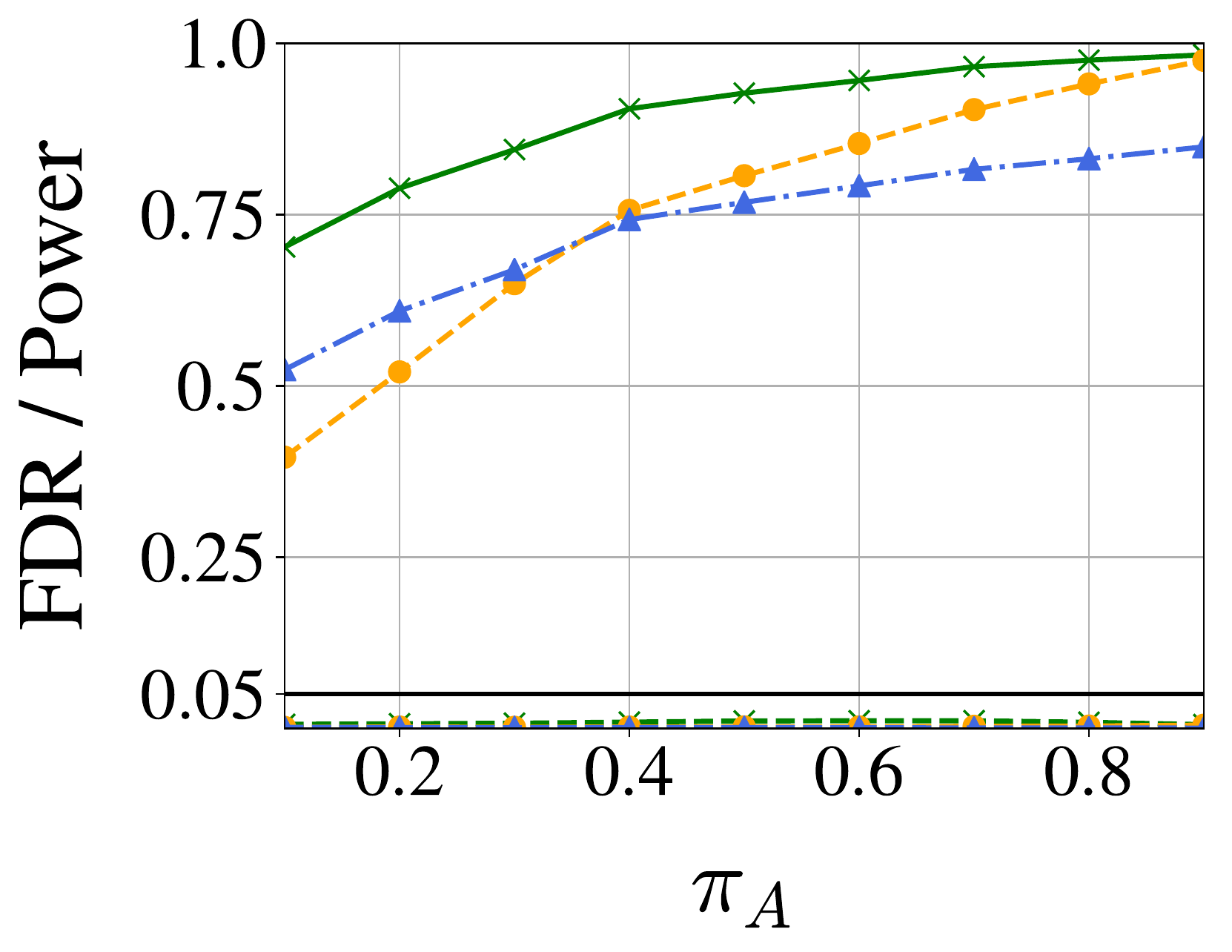}
	\end{subfigure}%
	\begin{subfigure}{0.4\textwidth}
		\centering
		\includegraphics[width=0.9\linewidth]{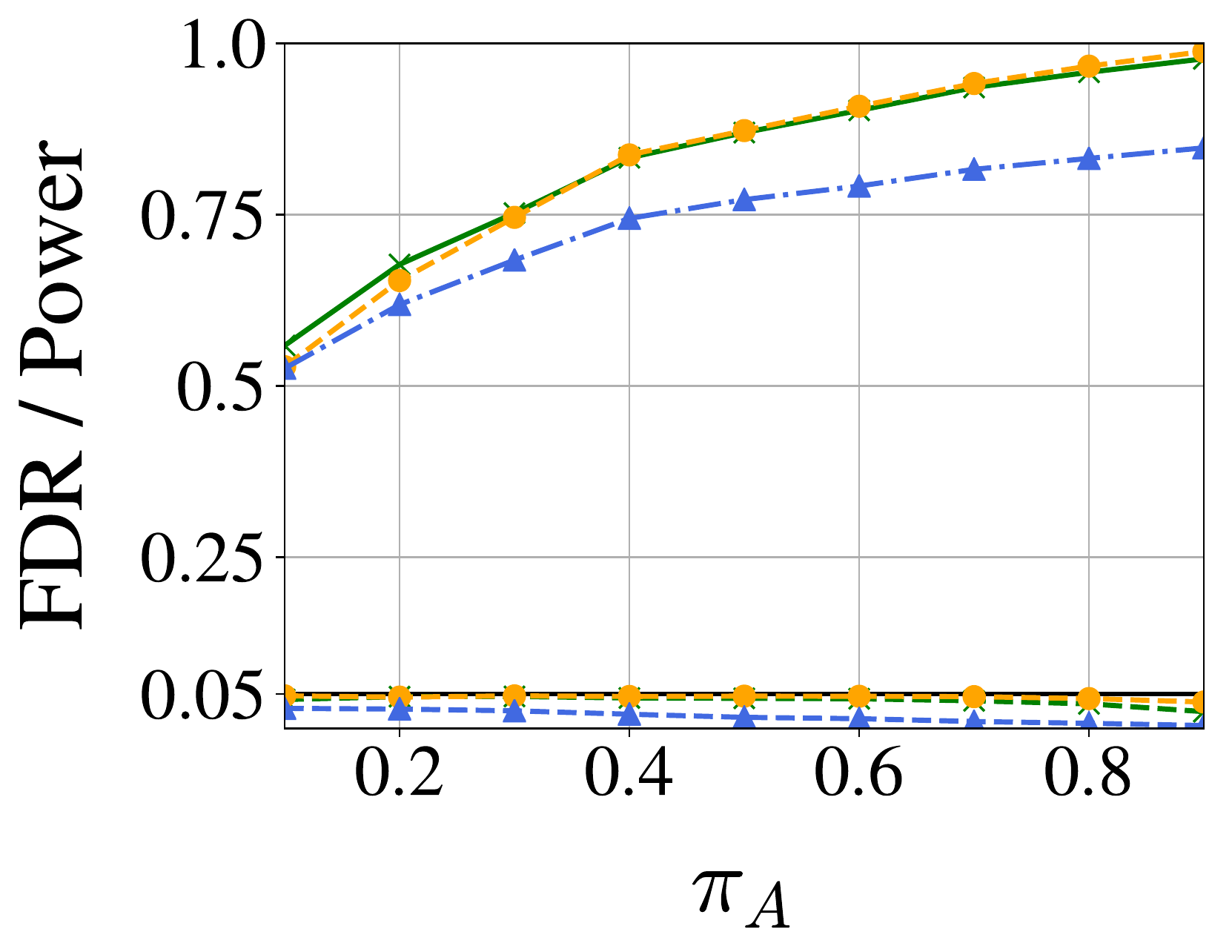}
	\end{subfigure}%
	\begin{subfigure}{0.1\textwidth}
		\centering
		\includegraphics[width=\linewidth]{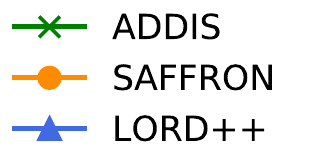}
	\end{subfigure}
	\caption{Statistical power and FDR versus fraction of non-null hypotheses $\pi_A$ for ADDIS, SAFFRON and LORD++ at target FDR level $\alpha= 0.05$ (solid black line). The curves above 0.05 line display the power of each methods versus $\pi_A$, while the lines below 0.05 display the FDR of each methods versus $\pi_A$. The experimental setting is described in \secref{simu}: we set $\mu_A = 3$ for both figures, but $\mu_N = -1$ for the left figure and $\mu_N = 0$ for the right figure (hence the left nulls are conservative, the right nulls are not). These figures show that (a) all considered methods do control the FDR at level 0.05, (b) SAFFRON sometimes loses its superiority over its nonadaptive variant LORD++ with conservative nulls (i.e. $\mu_N <0 $); and (c) ADDIS is more powerful than SAFFRON and LORD++ with conservative nulls, while loses almost nothing under settings with uniform nulls (i.e. $\mu_N = 0$).}\label{fig:powerloss}
\end{figure}

However, an important point is that SAFFRON is more powerful only when the $p$-values are exactly uniformly distributed under the null hypothesis. In practice, one frequently encounters \emph{conservative} nulls (see below), and in this case SAFFRON can have lower power than LORD++ (see \figref{powerloss}). 

 \paragraph{Uniformly conservative nulls.}
When performing hypothesis testing, we always assume that the $p$-value $P$ is \emph{valid}, which means that if the null hypothesis is true, we have $\PP{P\leq x}\leq x$ for all $x\in[0,1]$. Ideally, a $p$-value is exactly uniformly distributed, which means that the inequality holds with equality. However, we say a null $p$-value is \emph{conservative} if the inequality is strict, and often the nulls are \emph{uniformly conservative}, which means that
  under the null hypothesis, we have
	\begin{equation}\label{conserve-def}
	 \PPst{P/\tau \leq x}{P \leq \tau} \leq x \quad \text{ for all } x,\tau \in (0,1).
	\end{equation} 
\noindent
As an obvious first example, the $p$-values being exactly uniform (the ideal setting) is a special case. Indeed, for a uniform $U \sim U[0,1]$, if you know that $U$ is less than (say) $\tau=0.4$, then the conditional distribution of $U$ is just $U[0,0.4]$, which means that $U/0.4$ has a uniform distribution on $[0,1]$, and hence $\PPst{U/0.4 \leq x}{U \leq 0.4} \leq x$ for any $x \in (0,1)$. A mathematically equivalent definition of uniformly conservative nulls is that the CDF $F$ of a null $p$-value $P$ satisfies the following property:
\begin{equation}\label{conserve-def-equal}
F(\tau x) \leq x F(\tau), \ \ \ \ \text{ for all }\ 0\leq x, \tau \leq 1.
\end{equation}
\noindent
Hence, any null $p$-value with convex CDF is uniformly conservative. Particularly, when $F$ is differentiable, the convexity of $F$ is equivalent to its density $f$ being monotonically increasing. Here are two tangible examples of tests with uniformly conservative nulls: 
\begin{itemize}
\item A test of Gaussian mean: we test the null hypothesis $H_0 : \mu  \leq 0$ against the alternative $H_1: \mu > 0$; the observation is $Z \sim N(\mu,1)$ and the $p$-value is computed as $P = \Phi(-Z)$, where $\Phi$ is the standard Gaussian CDF. 
\item A test of Gaussian variance: we observe $Z \sim N(0,\sigma)$ and we wish to test the null hypothesis $H_0 : \sigma  \leq 1$ against the $H_1: \sigma > 1$ and the $p$-value is $P = 2\Phi(-|Z|)$. 
\end{itemize} 
It is easy to verify that, if the true $\mu $ in the first test is strictly smaller than zero, or the true $\sigma $ in the second test is strictly smaller than one, then the corresponding null $p$-values have monotonically increasing density, thus being uniformly conservative. 
More generally, \citet{zhao2018multiple} presented the following sufficient condition for a one-dimensional exponential family with true parameter $\theta$: when the true $\theta$ is strictly smaller than $\theta_0$, the uniformly most powerful (UMP) test of $H_0: \theta \leq \theta_0$ versus $H_1: \theta > \theta_0$ is uniformly conservative. Since the true underlying state of nature is rarely \emph{exactly} at the boundary of the null set (like $\mu=0$ or $\sigma=1$ or $\theta=\theta_0$ in the above examples), it is common in practice to encounter uniformly conservative nulls. In the context of A/B testing, this corresponds to testing $H_0: \mu_B \leq \mu_A$ against $H_1:\mu_B > \mu_A$, when in reality, $B$ (the new idea) is strictly worse than $A$ (the existing system), a very likely scenario.

\paragraph{Our contribution}
The main contribution of this paper is a new method called ADDIS (an ADaptive algorithm that DIScards conservative nulls), that compensates for the power loss of SAFFRON with conservative nulls. ADDIS is based on a new serial estimate of the false discovery proportion, having adaptivity to both fraction of nulls (like SAFFRON) and the conservativeness of nulls (unlike SAFFRON). As shown in \figref{powerloss}, ADDIS enjoys appreciable power increase over SAFFRON as well as LORD++  under settings with many conservative nulls, and rarely loses power when the nulls are exactly uniformly distributed (not conservative). 
Our work is motivated by recent work by \citet{zhao2018multiple} who study nonadaptive offline multiple testing problems with conservative nulls,  and ADDIS can be regarded as extending their work to both online and adaptive settings. The connection to the offline setting is that ADDIS effectively employs a ``discarding'' rule, which states we should discard (that is, not test) a hypothesis with $p$-value exceeding certain threshold. Beyond the online setting, we also incorporate this rule into several other existing FDR methods, and formally prove that the resulting new methods still control the FDR, while demonstrating numerically they have a consistent power advantage over the original methods. \figref{history} presents the relational chart of historical FDR control methods together with some of the new methods we proposed. As far as we know, we provide the first method that adapts to the conservativeness of nulls in the online setting.

\begin{minipage}{\columnwidth}\begin{center}
    \begin{tikzpicture}[node distance=1.1cm]
    
    \node (offline) [title] {Offline methods};    
    \node (BH) [noneoff, right of=offline, xshift = 1cm] {\scriptsize{BH \cite{BH95}}};
    \node (StBH) [adaoff, right of=BH, xshift = 3cm] {\scriptsize{Storey-BH \cite{storey2002direct}}};
    \node (D-StBH) [adadisoff, right of=StBH, xshift = 4cm] {\scriptsize{D-StBH (\secref{distorey}})};      
    
    \node (online) [title, below of=offline] {Online methods};
    \node (LORD) [none, right of=online, xshift = 1cm, align = center] {\scriptsize{LORD \cite{javanmard2017on}} \\ \scriptsize{LORD++ \cite{ramdas2017on}}};
    \node (SAFFRON) [ada, below of=StBH] {\scriptsize{SAFFRON \cite{ramdas2018saffron}}};
    \node (AI) [ada, below of=SAFFRON, yshift = 0.3cm] {\scriptsize{Alpha-Investing \cite{foster2008alpha}}};   
    \node (ADDIS) [adadis, below of=D-StBH] {\scriptsize{ADDIS (\secref{addis})}};
    
     \draw [arrow] (BH) --node[anchor=south, align = center]{\scriptsize{adaptivity}}(StBH);
      \draw [arrow] (StBH) --node[anchor=south, align = center]{\scriptsize{discarding}}(D-StBH);
    \draw [dotarrow] (BH) --node[anchor=west]{\scriptsize{online analog}}(LORD);
    \draw [dotarrow] (StBH) --node[anchor=west]{\scriptsize{online analog}}(SAFFRON);
    \draw [dotarrow] (D-StBH) --node[anchor=west]{\scriptsize{online analog}}(ADDIS);
    \draw [arrow] (LORD) -- node[anchor=south, align = center]{\scriptsize{adaptivity}}(SAFFRON);
    \draw [arrow] (SAFFRON) -- node[anchor=south, align = center]{\scriptsize{discarding}}(ADDIS);
    \draw [arrow] (SAFFRON) -- node[anchor=west]{\scriptsize{special case}}(AI);       

    \end{tikzpicture}
    
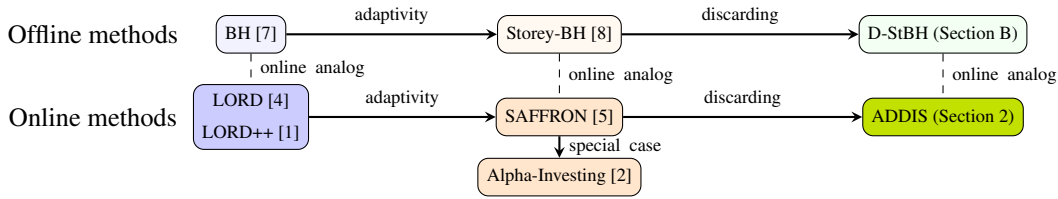
\captionof{figure}{Historical context: ADDIS generalizes SAFFRON, which generalizes Alpha-Investing and LORD++. Analogously, D-StBH (supplement) generalizes Storey-BH, which generalizes BH.}\label{fig:history}
    \end{center}\end{minipage}
    
\paragraph{Paper outline.}
 In \secref{addis}, we derive the ADDIS algorithm and state its guarantees (FDR and mFDR control), deferring proofs to the supplement. Specifically, in \secref{balance}, we discuss how to choose the hyperparameters in ADDIS to balance adaptivity and discarding for optimal power. \secref{simu} shows simulations which demonstrate the advantage of ADDIS over non-discarding or non-adaptive methods. We then generalize the ``discarding'' rule of ADDIS in \secref{general} and use it to obtain the ``discarding'' version of many other methods under various settings. We also show the error control with formal proofs for those variants in the supplement. Finally, we present a short summary in \secref{conclusion}. The code to reproduce all figures in the paper is included in the supplement.

\section{The ADDIS algorithm}\label{sec:addis}
Before deriving the ADDIS algorithm, it is useful to set up some notation. 
Recall that $P_j$ is the $p$-value for testing hypothesis $H_j$. For some sequences $\{\alpha_t\}_{t=1}^{\infty}$, $\{\tau_t\}_{t=1}^{\infty}$ and $\{\lambda_t\}_{t=1}^{\infty}$, where each term is in the range $[0,1]$, define the indicator random variables 
\[S_j = \ident\{P_j \leq \tau_j\}, \quad C_j = \ident\{ P_j \leq \lambda_j \}, \quad R_j = \ident\{ P_j \leq \alpha_j\}.\] 
They respectively answer the questions: ``\emph{was $H_j$ selected for testing? (or was it discarded?)}'', ``\emph{was $H_j$ a candidate for rejection?}'' and ``\emph{was $H_j$ rejected, yielding a discovery?}''. We call the sets \[
S(t) = \{j\in[t] : S_j =1\}, \quad C(t) = \{j\in[t]: C_j =1 \}, \quad R(t) = \{j\in[t]: R_j =1 \}
\]
as the ``selected (not discarded) set'', ``candidate set'' and ``rejection set'' after $t$ steps respectively. Similarly, we define $R_{1:t}= \{R_1,\dots,R_{t}\}$, $C_{1:t}= \{C_1,\dots,C_{t}\}$ and $S_{1:t} =\{S_1,\dots,S_{t}\}$.
In what follows in this section and the next section, we repeatedly encounter the filtration 
\[\F^t := \sigma(R_{1:t}, C_{1:t}, S_{1:t}).\] 
We insist that $\alpha_t$, $\lambda_t$ and $\tau_t$ are predictable, that is they are measurable with respect to $\Ft$. This means that $\alpha_t,\lambda_t,\tau_t$ are really mappings from $\{R_{1:t-1}, C_{1:t-1}, S_{1:t-1}\} \mapsto [0,1]$.

The presentation is cleanest if we assume that the $p$-values from the different hypotheses are independent (which would be the case if each A/B test was based on fresh data, for example). However, we can also prove mFDR control under a mild form of dependence: we call the null $p$-values \emph{conditionally uniformly conservative} if for any $t\in \nulls$, we have that
\begin{equation}\label{condition-conserve-def}
	\forall x,\tau \in (0,1),\ \PPst{P_t/\tau \leq x}{P_t \leq \tau, \Ft} \leq x.
\end{equation}
Note that the above condition is equivalent to the (marginally) uniformly conservative property \eqref{conserve-def} if the $p$-values are independent, and hence $P_t$ is independent of $\F^{t-1}$. For simplicity, we will refer this ``conditionally uniformly conservative'' property still as ``uniformly conservative''.  

\subsection{Deriving ADDIS algorithm}
Denote the (unknown) false discovery proportion by $\fdp \equiv \frac{|\nulls\cap\R(T)|}{|\R(T)|\vee 1}$. As mentioned in \cite{ramdas2018saffron}, one can control the FDR at any time $t$ by instead controlling an oracle estimate of the FDP, given by
\begin{equation}\label{fdp*}
	\fdp^{*}(t) \defn \frac{\sum_{j \leq t, j\in\nulls}\alpha_j}{|R(t)|\vee 1}.
\end{equation}
This means that if we can keep $\fdp^{*}(t) \leq \alpha$ at all times $t$, then we can prove that $\fdr(t) \leq \alpha$ at all times $t$.  Since the set of nulls $\nulls$ is unknown, LORD++ \cite{ramdas2017on} is based on the simple upper bound of $\fdp^*(t)$, defined as $\widehat{\fdp}_\textnormal{LORD++}(t)$, and SAFFRON \cite{ramdas2018saffron} is based on a more nuanced adaptive bound on $\fdp^*(t)$, defined as $\widehat{\FDP}_{\textnormal{SAFFRON}}(t)$, obtained by choosing a predictable sequence $\{\lambda_j\}_{j=1}^{\infty}$; where
\begin{equation}\label{fdpsaffron}
\widehat{\fdp}_\textnormal{LORD++}(t) \defn \frac{\sum_{j \leq t}\alpha_j}{|R(t)|\vee 1}, \ \ \ \ \widehat{\FDP}_{\textnormal{SAFFRON}}(t) \defn \frac{\sum_{j \leq t}\alpha_j\frac{\ident\{P_j > \lambda_j \}}{1-\lambda_j}}{|R(t)|\vee 1}.
\end{equation}
It is easy to fix $\alpha_1 < \alpha$, and then update $\alpha_2,\alpha_3,\dots$ in an online fashion to maintain the invariant $\fdphat_\textnormal{LORD++}(t) \leq \alpha$ at all times, which the authors prove suffices for FDR control, while it is also proved that keeping $\widehat{\FDP}_{\textnormal{SAFFRON}}(t) \leq \alpha$ at all times suffices for FDR control at any time. However, we expect $\widehat{\FDP}_{\textnormal{SAFFRON}}(t)$ to be closer
\footnote{To see this intuitively, consider the case when (a) $\lambda_j \equiv 1/2$ for all $j$, (b) there is a significant fraction of non-nulls, and the non-null $p$-values are all smaller than 1/2 (strong signal), and (c) the null $p$-values are exactly uniformly distributed. Then, $\frac{\ident\{1/2 < P_j \}}{1/2}$ evaluates to 0 for every non-null, and equals one for every null in expectation. Thus, in this case, $\EE{\sum_{j \leq t}\alpha_j\frac{\ident\{\lambda_j < P_j \}}{1-\lambda_j}} = \EE{\sum_{j \leq t, j\in\nulls}\alpha_j} \ll \EE{\sum_{j \leq t}\alpha_j}$.}  to $\fdp^*(t)$ than $\widehat{\FDP}_{\textnormal{LORD++}}(t)$, and since SAFFRON better uses its FDR budget, it is usually more powerful than LORD++. SAFFRON is called an ``adaptive'' algorithm, because it is the online analog of the Storey-BH procedure \cite{storey2002direct}, which adapts to the proportion of nulls in the offline setting.

However, in the case when there are many conservative null $p$-values (whose distribution is stochastically larger than uniform), many terms in  $\{\frac{\ident\{\lambda_j < P_j \}}{1-\lambda_j}: j\in \nulls\}$ may have  expectations much larger than one, making $\widehat{\FDP}_{\textnormal{SAFFRON}}(t)$ an overly conservative estimator of $\fdp^{*}(t)$, and thus causing a loss in power. In order to fix this, we propose a new empirical estimator of $\fdp^{*}(t)$. We pick two predictable sequences $\{\lambda_j\}_{j=1}^{\infty}$ and $\{\tau_j\}_{j=1}^{\infty}$ such that $\lambda_j < \tau_j$ for all $j$, for example setting $\lambda_j=1/4, \tau_j=1/2$ for all $j$, and define
\begin{equation}\label{fdpaddis}
	\widehat{\FDP}_{\textnormal{ADDIS}}(t) \defn \frac{\sum_{j \leq t}\alpha_j\frac{\ident\{\lambda_j < P_j \leq \tau_j\}}{\tau_j-\lambda_j}}{|R(t)| \vee 1} ~\stackrel{\theta_j \defn \frac{\lambda_j}{\tau_j}}{\equiv}~ \frac{\sum_{j \leq t}\alpha_j\frac{\ident\{ P_j \leq \tau_j\} \ident\{P_j/\tau_j > \theta_j\}}{\tau_j(1-\theta_j)}}{|R(t)|\vee 1}
\end{equation}
\noindent
With many conservative nulls, the claim that ADDIS is more powerful than SAFFRON, is based on the idea that the numerator of $\widehat{\FDP}_{\textnormal{ADDIS}}(t)$ is a much tighter estimator of $\sum_{j \leq t, j\in\nulls} \alpha_j$, compared with that of $\widehat{\FDP}_{\textnormal{SAFFRON}}(t)$. In order to see why this is true, we provide the following lemma.
\begin{lemma}\label{lem:estimates}
	If a null $p$-value $P$ has a differentiable convex CDF, then for any constants $a,b \in (0,1)$, we have 
	\begin{equation}\label{estimates}
		\frac{\PP{ab < P \leq b}}{b(1-a)} \leq \frac{\PP{ P > a }}{(1-a)}.
	\end{equation}
\end{lemma}
\noindent
The proof of \lemref{estimates} is presented in \secref{pfestimates}. 
Recalling definition \eqref{conserve-def-equal}, \lemref{estimates} implies that for some uniformly conservative nulls, our estimator $\widehat{\FDP}_{\textnormal{ADDIS}}$ will be tighter than $\widehat{\FDP}_{\textnormal{SAFFRON}}$ in expectation, and thus an algorithm based on keeping $\widehat{\FDP}_{\textnormal{ADDIS}} \leq \alpha$ is expected to have higher power. 

\paragraph{ADDIS algorithm} We now present the general ADDIS algorithm. Given user-defined sequences $\{\lambda_j\}_{j=1}^{\infty}$ and $\{\tau_j\}_{j=1}^{\infty}$ as described previously, we call an online FDR algorithm as an instance of the ``ADDIS algorithm'' if it updates $\alpha_t$ in a way such that it maintains the invariant $\fdphat_{\textnormal{ADDIS}}(t) \leq \alpha $. 
We also enforce the constraint that $\tau_t > \lambda_t \geq \alpha_t$ for all $t$, which is needed for correctness of the proof of FDR control. This is not a major restriction since we often choose $\alpha = 0.05$, and the algorithms set $\alpha_t \leq \alpha$, in which case $\tau_t > \lambda_t \geq 0.05$ easily satisfies the needed constraint. Now, the main nontrivial question is how to ensure the  invariant in a fully online fashion. We address this  by providing an explicit instance of ADDIS algorithm, called ADDIS$^{*}$ (Algorithm~\ref{addisalgo}), in the following section.
From the form of the invariant $\fdphat_{\textnormal{ADDIS}}(t) \leq \alpha $, we observe that any $p$-value $P_j$ that is bigger than $\tau_j$ has no influence on the invariant, as if it never existed in the sequence at all. This reveals that ADDIS effectively implements a ``discarding" rule: it discards $p$-values exceeded a certain threshold. If the $p$-value is not discarded, then $P_j/\tau_j$ is a valid $p$-value and we resort to using adaptivity like \eqref{fdpsaffron}.

\subsection{ADDIS$^{*}$: an instance of ADDIS algorithm using constant $\lambda$ and $\tau$}
Here we present an instance of ADDIS algorithm, with choice of  $\lambda_j \equiv \lambda$ and $\tau_j \equiv \tau$ for all $j$. (We consider constant $\lambda$ and $\tau$ for simplicity, but these can be replaced by $\lambda_j$ and $\tau_j$ at time $j$.)

\begin{algorithm}[H]\label{addisalgo}
	\KwIn{FDR level $\alpha$, discarding threshold $\tau \in (0,1]$, candidate threshold $\lambda \in [0,\tau)$, sequence $\{\gamma_j \}_{j=0}^{\infty}$ which is nonnegative, nonincreasing and sums to one, initial wealth $W_0 \leq \alpha$.}
	\For{$t=1, 2, \dots$}{
		Reject the $t$-th null hypothesis if  $P_t\leq \alpha_t$, where $\alpha_t \defn \min\{\lambda, \widehat{\alpha}_{t}\}, \label{constrain} $ and \\
		$\widehat{\alpha}_{t}  \defn (\tau-\lambda) \left(W_0\gamma_{S^t - C_{0+}} +(\alpha-W_0)\gamma_{S^t-\kappa_1^{*}-C_{1+}}  + \alpha\sum_{j\geq 2} \gamma_{S^t - \kappa_j^{*}-C_{j+}}\right).$\\
		Here, $\ \ S^{t} = \sum_{i< t} \ident\{ P_i \leq \tau\} $, $ \ \ \ \ C_{j+} = \sum_{i=\kappa_j+1}^{t-1}\ident\{ P_i \leq \lambda \}$, \\
		$ \ \ \ \  \ \ \ \ \ \ \ \ \kappa_j = \min \{i\in [t-1]: \sum_{k\leq i} \ident\{ P_k \leq \alpha_k\} \geq j\}$, $ \ \ \ \  \kappa_j^{*} = \sum_{i \leq \kappa_j} \ident\{ P_i \leq \tau\} $.
	}
	\caption{The $\textnormal{ADDIS}^{*}$ algorithm}
\end{algorithm}

 In \secref{verify}, we verify that $\alpha_t$ is a monotonic function of the past\footnote{We say that a function $f_t(R_{1:t-1}, C_{1:t-1}, S_{1:t-1}) : \{0,1\}^{3(t-1)} \to [0,1]$ is a monotonic function of the past, if $f_t$ is coordinatewise nondecreasing  in $R_{i}$ and $C_{i}$, and is coordinatewise nonincreasing in $S_{i}$. This is a generalization of the monotonicity of SAFFRON \cite{ramdas2018saffron}, which is recovered by setting $S_i=1$ for all $i$, that is we never discard any $p$-value.}\label{mono}. In \secref{addisalgoeq}, we present Algorithm~\ref{addisalgoeq}, which is an equivalent version of the above ADDIS$^{*}$ algorithm, but it explicitly discards $p$-values larger than $\tau$, thus justifying our use of the term ``discarding'' throughout this paper. Note that if we choose $\lambda \geq \alpha$, then the constraint $\alpha_t \defn \min{\{\lambda, \widehat{\alpha}_t\}}$ is vacuous and reduces to $\alpha_t \defn \widehat{\alpha}_t$, because $\widehat{\alpha}_t \leq \alpha$ by construction. 
 The power of ADDIS varies with $\lambda$ and $\tau$, as discussed further in \secref{balance}. 

\subsection{Error control of ADDIS algorithm}\label{sec:fdraddis}
Here we present error control guarantees for ADDIS, and defer proofs to \secref{pfaddisthm} and \secref{pfstopping}. 

\begin{theorem}\label{thm:addisthm}
	If the null $p$-values are uniformly conservative \eqref{condition-conserve-def}, and suppose we choose $\alpha_j,\lambda_j$ and $\tau_j$ such that $\tau_j > \lambda_j  \geq \alpha_j$ for each $j \in \mathbb{N}$, then we have:\\
	\-\hspace{0.5cm}
	(a) any algorithm with $\fdphat_{\textnormal{ADDIS}}(t) \leq \alpha$ for all $t \in \N$ also enjoys $\mfdr(t) \leq \alpha$ for all $t \in \N$.\\
	If we additionally assume that the null $p$-values are independent of each other and of
	the non-nulls, and always choose $\alpha_t$, $\lambda_t$ and $1-\tau_t$ to be monotonic functions of the past for all $t$, then we additionally have:\\
	\-\hspace{0.5cm} (b) any algorithm with $\fdphat_{\textnormal{ADDIS}}(t) \leq \alpha$ for all $t
		\in \N$
		 also enjoys $\fdr(t) \leq \alpha$ for all $t \in \N$.\\
As an immediate corollary, any ADDIS algorithm enjoys $\mfdr$ control, and ADDIS$^{*}$ (Algorithm~\ref{addisalgo}) additionally enjoys $\fdr$ control since it is a monotonic rule. 
\end{theorem}
\noindent
The above result only holds for nonrandom times. Below, we also show that any ADDIS algorithm controls mFDR at any stopping time with finite expectation.

\begin{theorem}\label{thm:stopping}
	Assume that the null $p$-values are uniformly conservative, and that $\min_j\{\tau_j-\lambda_j\}>\epsilon$ for some $\epsilon>0$. Then, for any stopping time $T_\textnormal{stop}$ with finite expectation, any algorithm that maintains the invariant $\fdphat_{\textnormal{ADDIS}}(t) \leq \alpha$ for all $t$ enjoys $\mfdr(T_\textnormal{stop}) \leq \alpha.$
\end{theorem}	

Once more, the conditions for the theorem are not restrictive because the sequences $\{\lambda_j\}_{j=1}^{\infty}$ and $\{\tau_j\}_{j=1}^{\infty}$ are user-chosen, and $\lambda_j=1/4, \tau_j=1/2$ is a reasonable default choice, as we justify next.

\subsection{Choosing $\tau$ and $\lambda$ to balance adaptivity and discarding}\label{sec:balance}
\begin{figure}[H]
	\centering
	\includegraphics[width=0.34\linewidth]{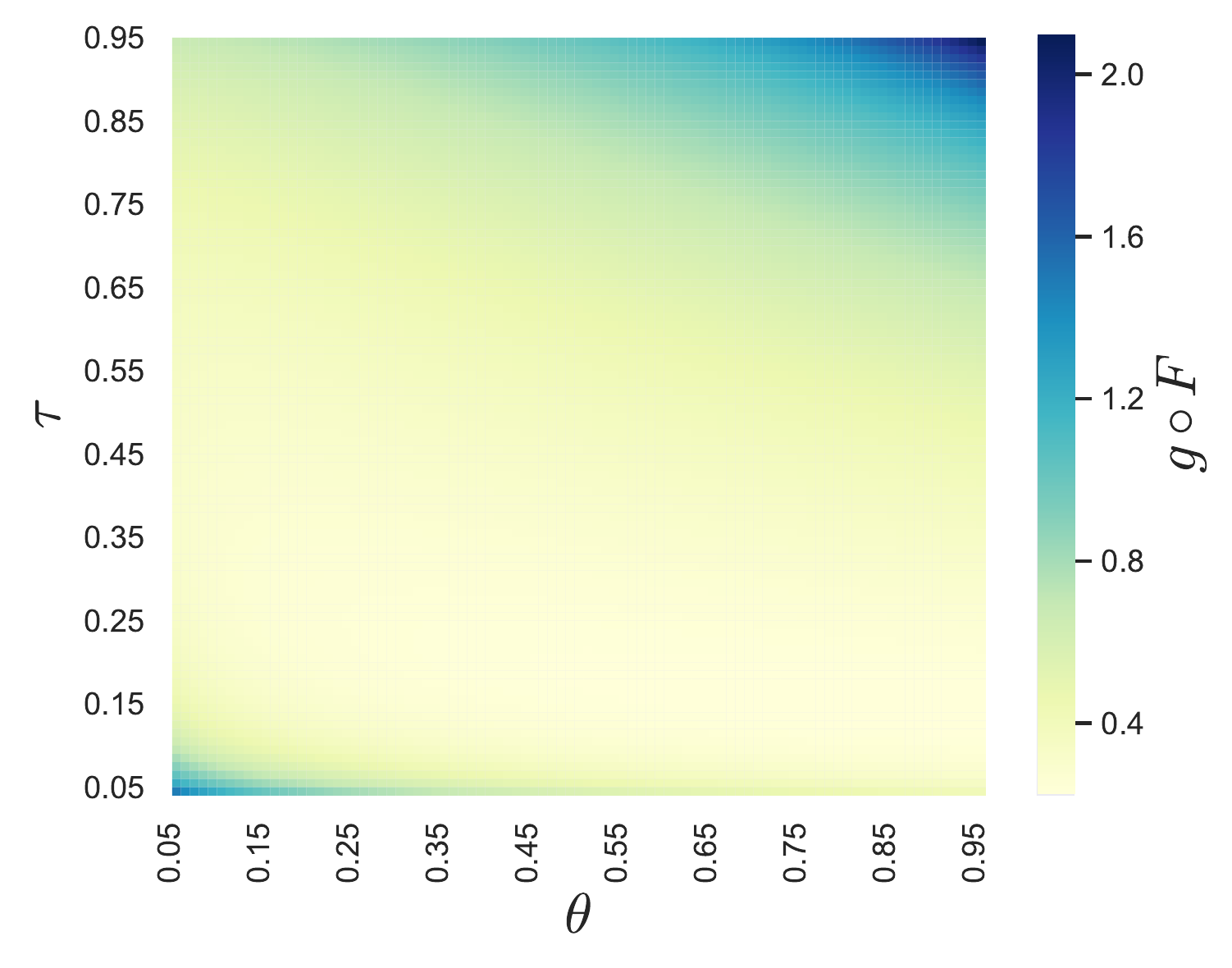}
	\includegraphics[width=0.34\linewidth]{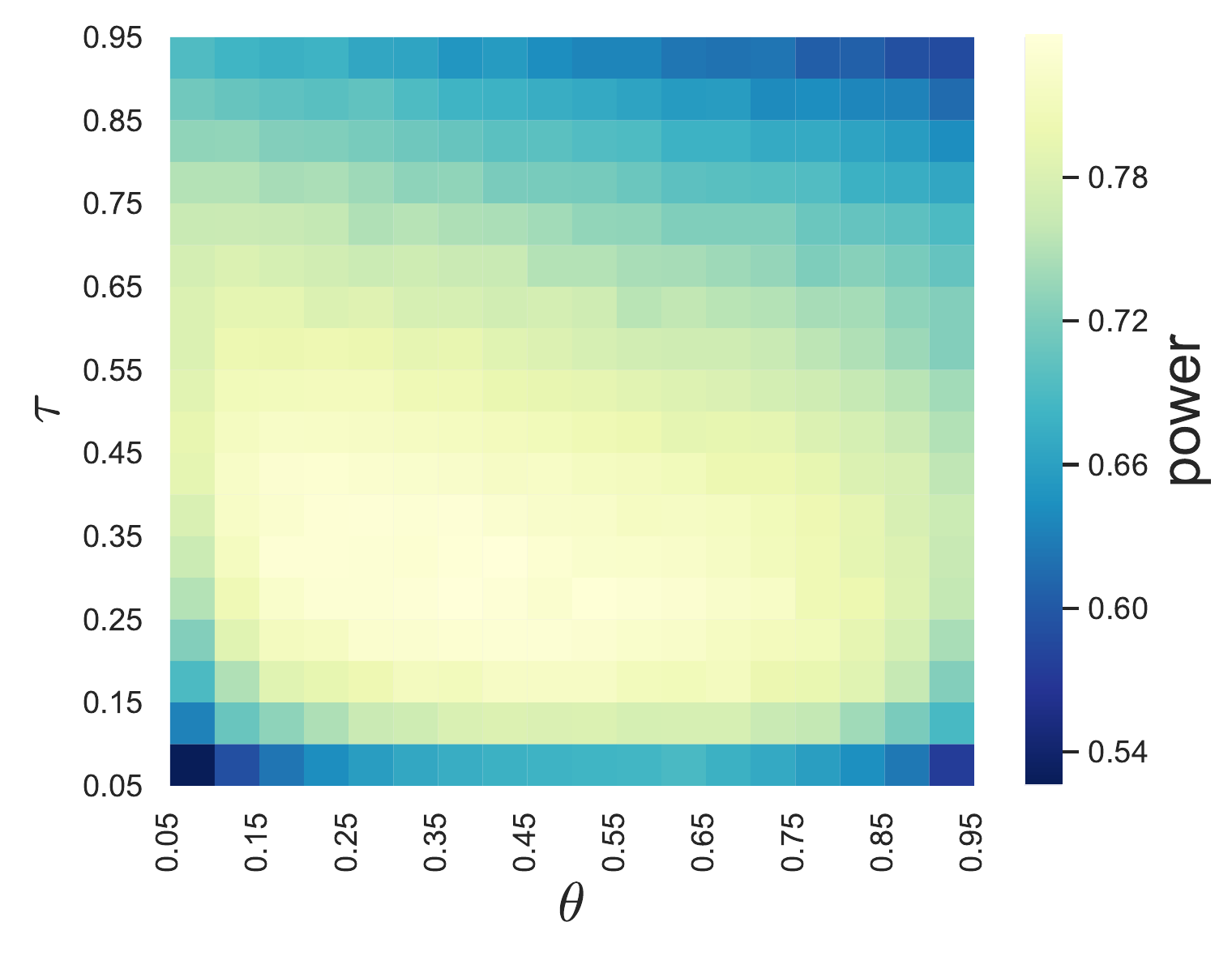}
	\caption{The left figure shows the heatmap of function $g\circ F$, where $F$ is the CDF of $p$-values drawn as described in \secref{simu} with $\mu_N = -1, \mu_A = 3, \pi_A = 0.2$. The right figure is the empirical power of ADDIS$^*$ versus different choice of $\theta$ and $\tau$, with $p$-values drawn from $F$. The figures are basically of the same pattern, with similar optimal values of $\theta$ and $\tau$. }\label{fig:tradeoff}
\end{figure}

As we mentioned before, the power of our $\textnormal{ADDIS}^{*}$ algorithm is closely related to the hyper-parameters $\lambda$ and  $\tau$. In fact, there is also an interaction between the hyper-parameters $\lambda$ and $\tau$, which means that one cannot decouple the effect of each on power. One can see this interaction clearly in \figref{tradeoff} which displays a trade off between adaptivity ($\lambda$) and discarding ($\tau$). Indeed, the right sub-figure displays a ``sweet spot'' for choosing $\lambda,\tau$, which should neither be too large nor too small.

Ideally, one would hope that there exists some universally optimal choice of $\lambda,\tau$ that yields maximum power. Unfortunately, the relationship between power and these parameters changes with the underlying distribution of the null and alternate $p$-values, as well as their relative frequency. Therefore, below, we only provide a heuristic argument about how to tune these parameters for $\textnormal{ADDIS}^{*}$.

Recall that the $\textnormal{ADDIS}^{*}$ algorithm is derived by tracking the empirical estimator $\fdphat_{\textnormal{ADDIS}}$ \eqref{fdpaddis} with fixed $\lambda$ and $\tau$, and keeping it bounded by $\alpha$ over time. Since  $\fdphat_{\textnormal{ADDIS}}$ serves as an estimate of the oracle $\fdp^{*}$ \eqref{fdp*}, it is natural to expect higher power with a more refined (i.e. tighter)  estimator $\fdphat_{\textnormal{ADDIS}}$.  One simple way to choose $\lambda$ and $\tau$ is to minimize the expectation of the indicator term in the estimator. Specifically, if the CDF of all $p$-values is $F$, then an oracle would choose $\lambda, \tau$ as 
\begin{equation}\label{minfdp}
	(\lambda^{*}, \tau^{*}) \in \arg \min_{\lambda < \tau \in(0,1)}{\frac{F(\tau) - F(\lambda)}{\tau - \lambda}}.
\end{equation}
 In order to remove the constraints between the two variables,  we again define $\theta = \lambda/\tau$, then the optimization problem \eqref{minfdp} is equivalent to
\begin{equation}
    (\theta^{*}, \tau^{*}) \in \arg \min_{\theta , \tau \in(0,1)}{\frac{F(\tau) - F(\theta\tau)}{\tau(1-\theta)}} \equiv (g\circ F) (\theta, \tau).
\end{equation}
We provide some empirical evidence to show the quality of the above proposal. The left subfigure in \figref{tradeoff} shows the heatmap of  $(g\circ F)$ and the right one shows the empirical power of $\textnormal{ADDIS}^{*}$ with $p$-values generate from $F$ versus different $\theta$ and $\tau$ (the left is simply evaluating a function, the right requires repeated simulation). The same pattern is consistent across other reasonable choices of $F$, as shown in \secref{secgf}. We can see that the two subfigures in \figref{tradeoff} show basically the same pattern, with similar optimal choices of parameters $\theta$ and $\tau$. Therefore, we suggest choosing $\lambda$ and $\tau$ as defined in \eqref{minfdp}, if prior knowledge of $F$ is available; otherwise it seems like $\theta \in [0.25, 0.75]$ and $\tau \in [0.15, 0.55]$ are safe choices, and for simplicity we use $\tau = \theta = 0.5$ as defaults, that is $\tau = 0.5, \lambda = 0.25$, in similar experimental settings. We leave the study of time-varying $\lambda_j$ and $\tau_j$ as future work.

\section{Numerical experiments}\label{sec:simu}
In this section, we numerically compare the performance of ADDIS against the previous state-of-the-art algorithm SAFFRON \cite{ramdas2018saffron}, and other well-studied algorithms like LORD++ \cite{javanmard2017on}, LOND \cite{javanmard2015on} and Alpha-investing \cite{foster2008alpha}. Specifically, we use $\textnormal{ADDIS}^{*}$ defined in Algorithm~\ref{addisalgo} as the representative of our ADDIS algorithm. Though as discussed in \secref{balance}, there is no universally optimal constants, given the minimal nature of our assumptions, we will use some reasonable default choices in the numerical studies to have a glance at the advantage of ADDIS algorithm. The constants $\lambda = 0.25$, $\tau = 0.5$ and sequence $\{\gamma_j\}^{\infty}_{j=0}$ with $\gamma_j \propto 1/(j+1)^{-1.6}$ were found to be particularly successful, thus are our default choices for hyperparameters in $\textnormal{ADDIS}^{*}$. We choose the infinite constant sequence $\gamma_j \propto \frac{1}{(j+1)^{1.6}}$, and $\lambda=0.5$ for SAFFRON, which yielded its best performance. We use $\gamma_j \propto \frac{\log{((j+1) \land 2)}}{(j+1)e^{\sqrt{\log{(j+1)}}}}$ for LORD++ and LOND, which is shown to maximize its power in the Gaussian setting \cite{javanmard2017on}. The proportionality constant of $\{\gamma_j\}_{i=0}^{\infty}$ is determined so that the sequence $\{\gamma_j\}_{i=0}^{\infty}$ sums to one. 

\begin{figure}[H]

	\begin{subfigure}{0.32\textwidth}
		\centering
		\includegraphics[width=\linewidth]{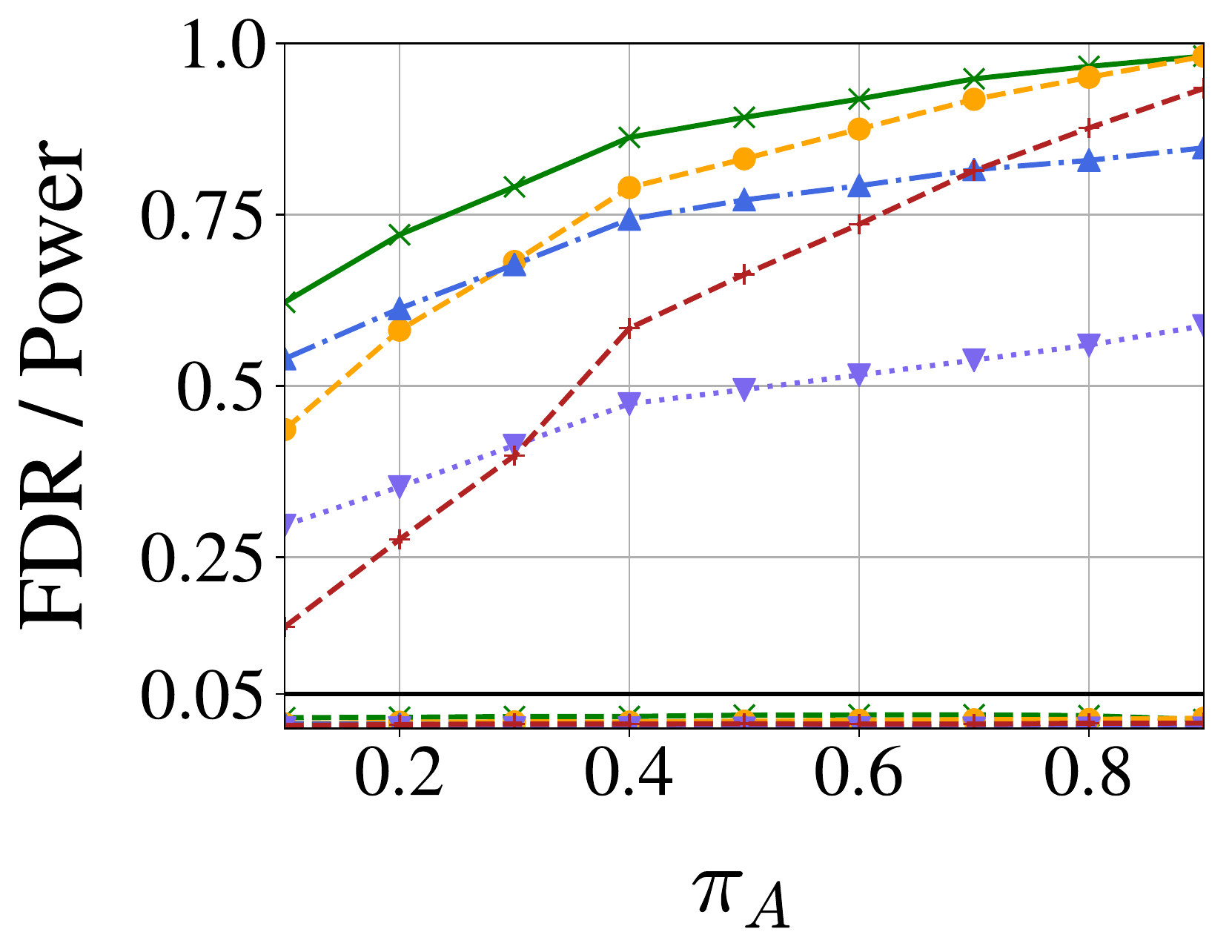}
		\caption{}
	\end{subfigure}%
	\begin{subfigure}{0.32\textwidth}
		\centering
		\includegraphics[width=\linewidth]{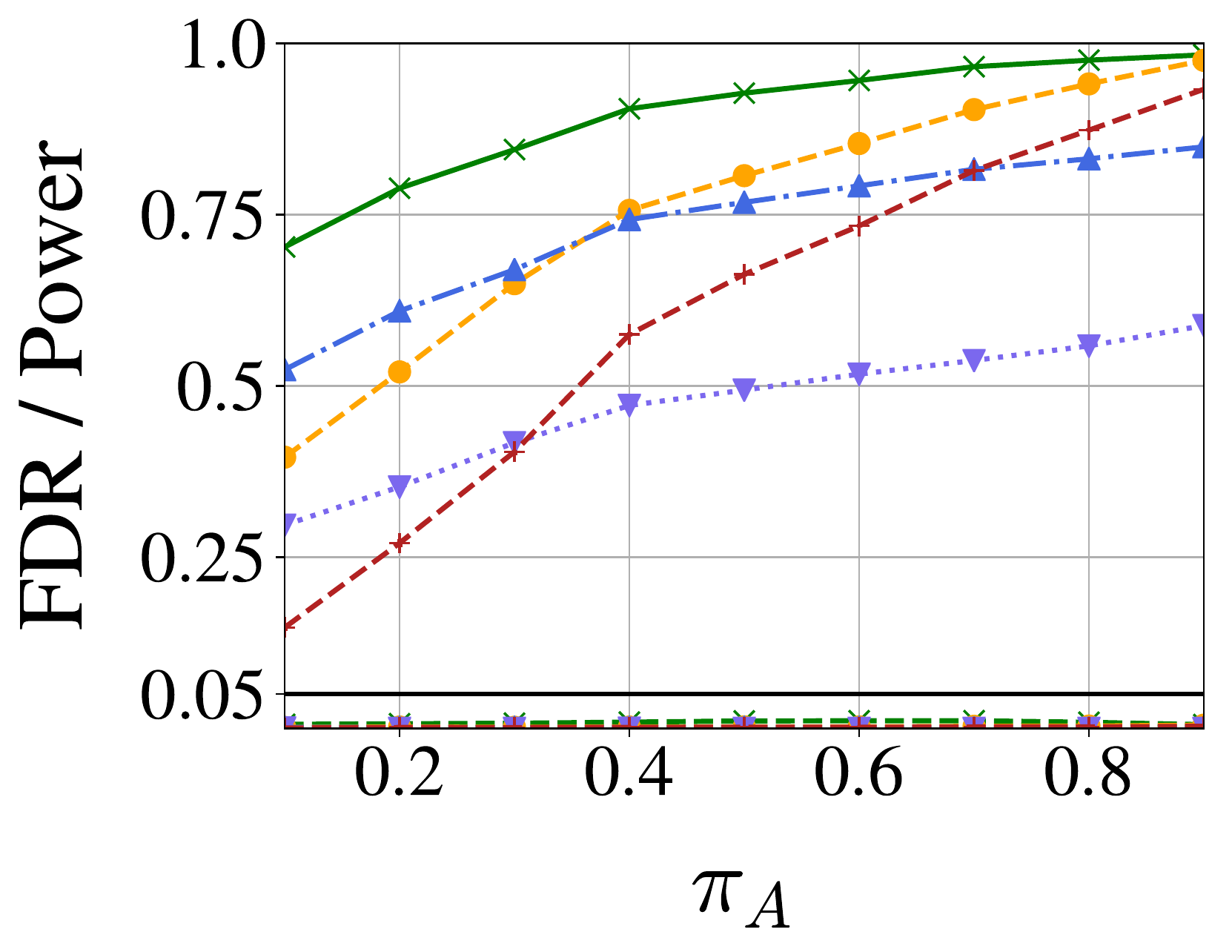}
		\caption{}
	\end{subfigure}	%
	\begin{subfigure}{0.32\textwidth}
		\centering
		\includegraphics[width=\linewidth]{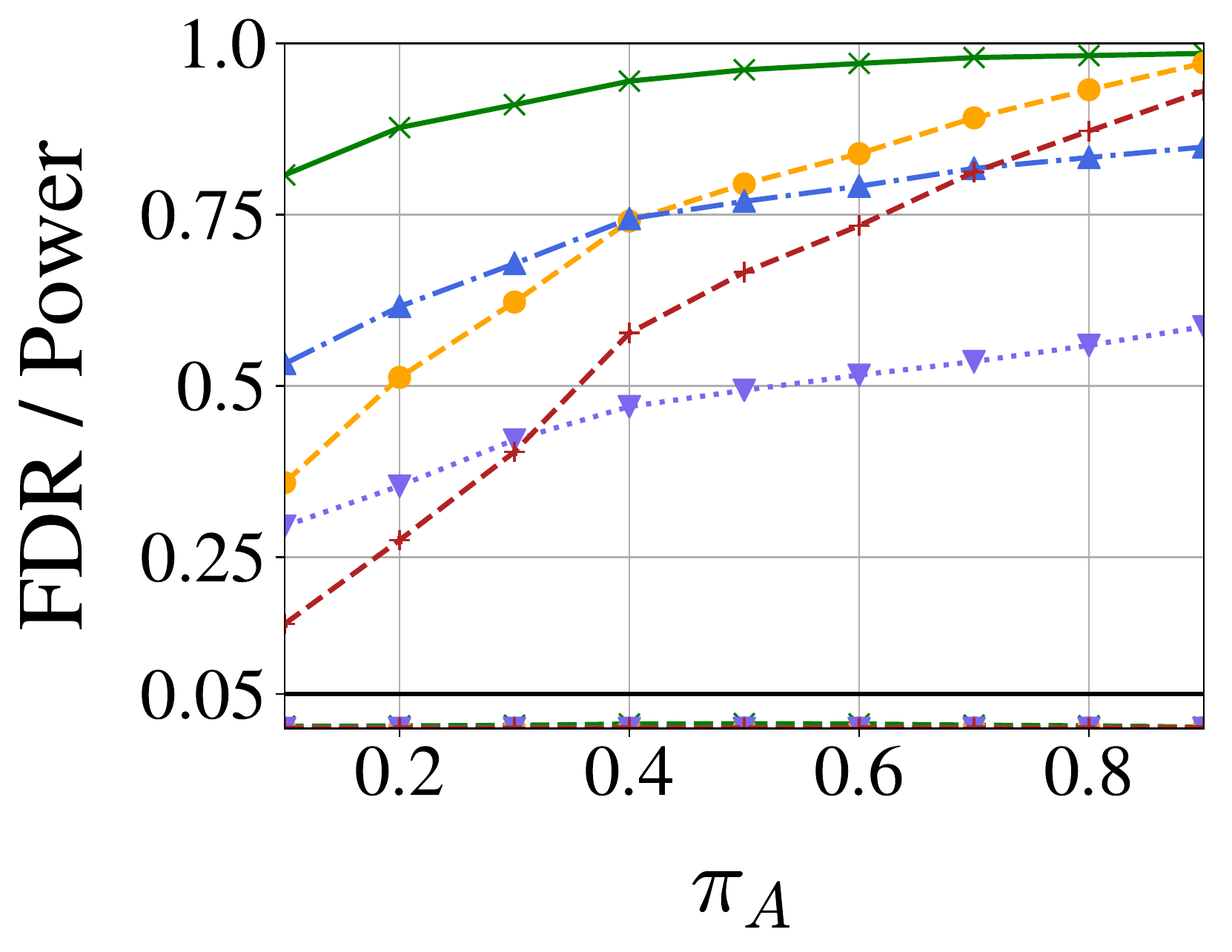}
		\caption{}
	\end{subfigure}\\

	\begin{subfigure}{0.32\textwidth}
		\centering
		\includegraphics[width=\linewidth]{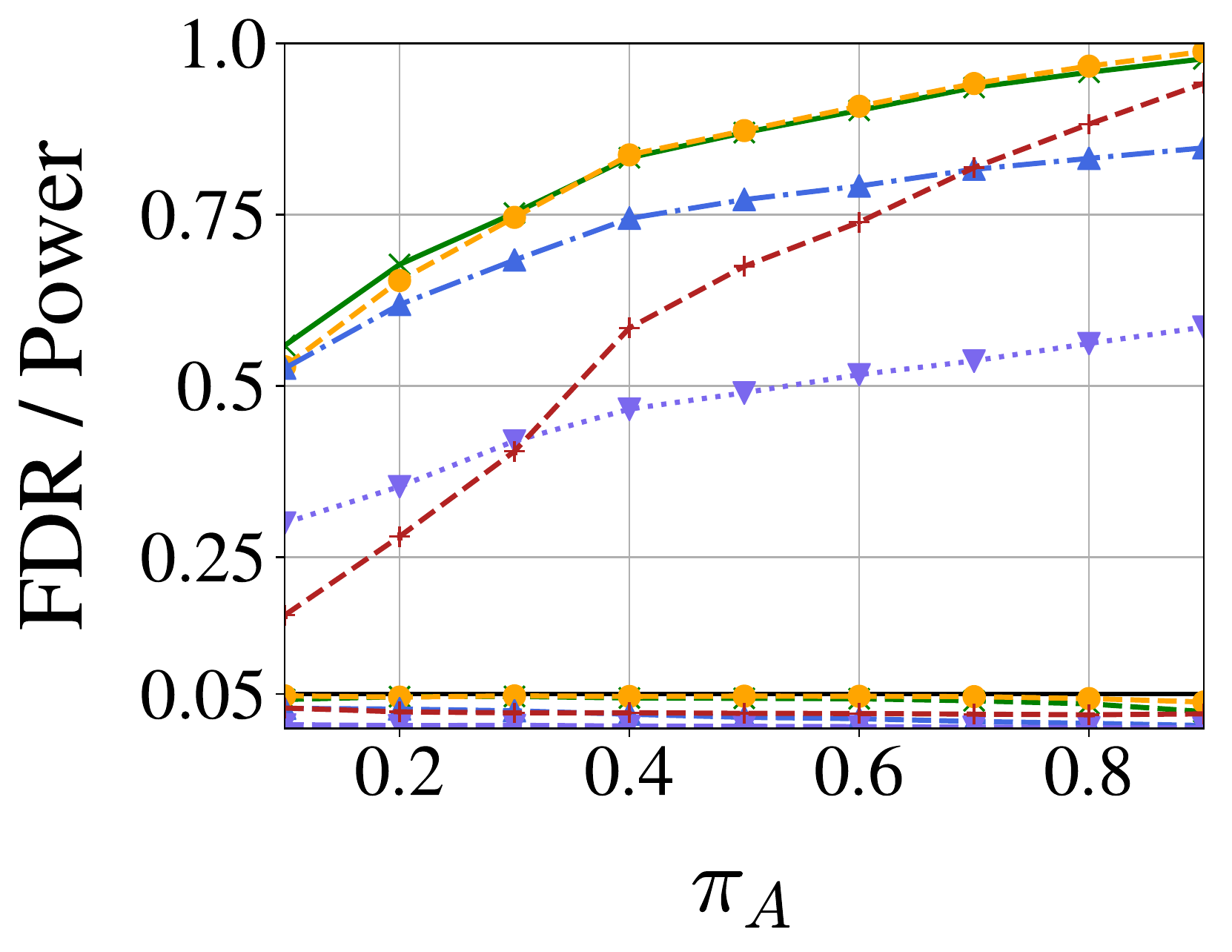}
		\caption{}
	\end{subfigure}%
	\begin{subfigure}{0.32\textwidth}
		\centering
		\includegraphics[width=\linewidth]{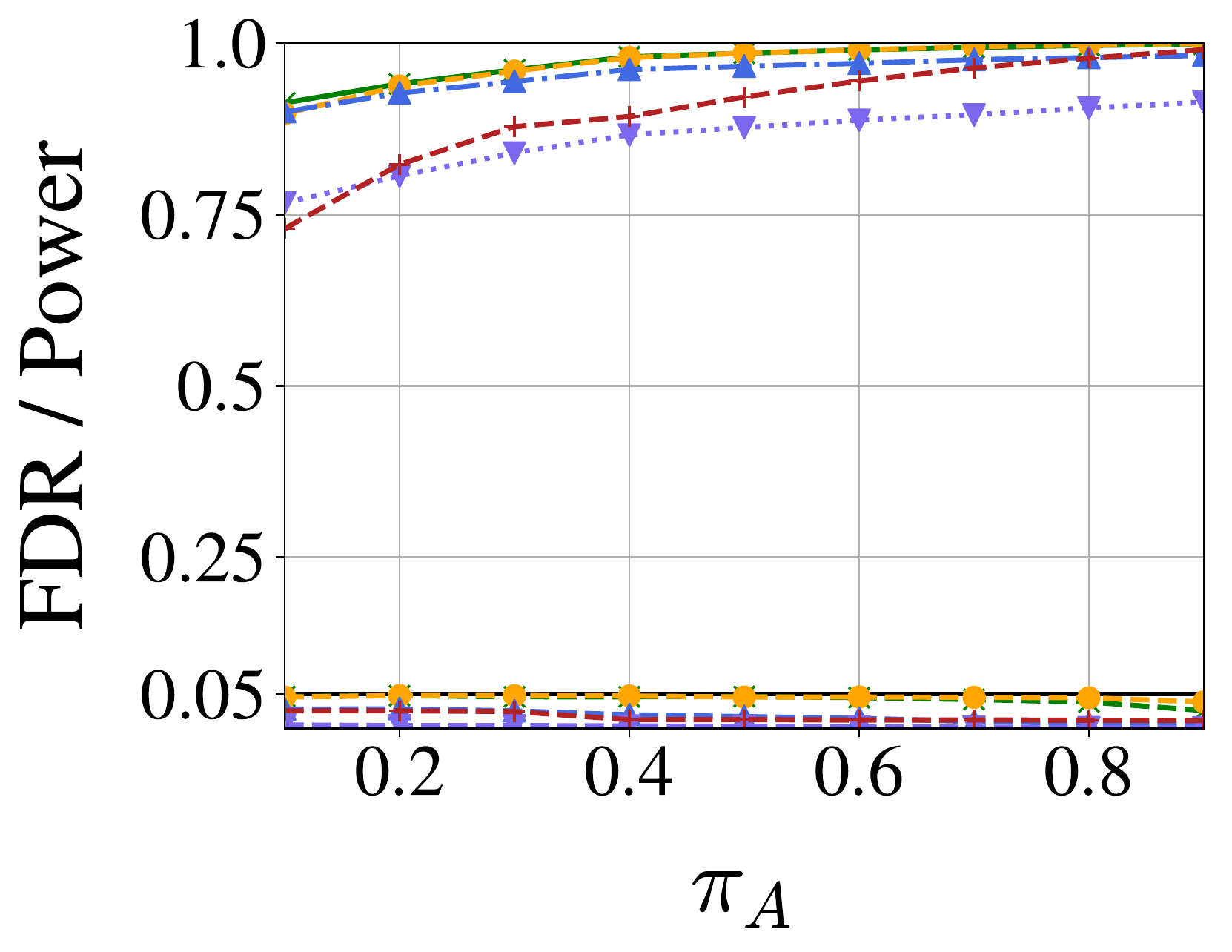}
		\caption{}
	\end{subfigure}
	\begin{subfigure}{0.1\textwidth}
		\centering
		\includegraphics[width=1.5\linewidth]{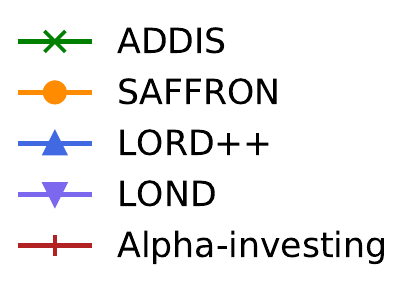}
		\caption*{}
	\end{subfigure}\\
	\caption{Statistical power and FDR versus fraction of non-null hypotheses $\pi_A$ for ADDIS, SAFFRON, LORD++, LOND, and Alpha-investing at target FDR level $\alpha$ = 0.05 (solid black line). The lines above the solid black line are the power of each methods versus $\pi_A$, and the lines below are the FDR of each methods versus $\pi_A$. The $p$-values are drawn using the Gaussian model as described in the text, while we set $\mu_N = -0.5$ in plot (a), $\mu_N = -1$ in plot (b), $\mu_N = -1.5$ in plot (c), and $\mu_N = 0$ in plots (d) and (e). And we set $\mu_A = 3$ in plots (a, b, c, d), $\mu_A = 4$ in plot (e). These plots show that (1) FDR is under control for all methods in all settings; (2) ADDIS enjoys appreciable power increase as compared to all the other four methods; (3) the more conservative the nulls are (the more negative $\mu_N$ is), the more significant the power increase of ADDIS is; (4) ADDIS matches SAFFRON and remains the best in the setting with uniform (not conservative) nulls. }\label{fig:ad}
\end{figure}

We consider the standard experimental setup of testing Gaussian means, with $M = 1000$ hypotheses. More precisely, for each index $i \in \{1,2,\dots,M\}$, the null hypotheses take the form $H_i: \mu_i \leq 0$, which are being tested against the alternative $H_{iA}: \mu_i > 0$. The observations are independent Gaussians $Z_i \sim N(\mu_i,1)$, where $\mu_i \equiv \mu_N \leq 0$ with probability $1-\pi_A$ and $\mu_i \equiv \mu_A>0$ with probability $\pi_A$. The one-sided $p$-values are computed as $P_i = \Phi(-Z_i)$, which are uniformly conservative if $\mu_N <0$ as discussed in the introduction (and the lower $\mu_N$ is, the more conservative the $p$-value). In the rest of this section, for each algorithm, we use target FDR $\alpha = 0.05$ and estimate the empirical FDR and power by averaging over 200 independent trials. \figref{ad} shows that ADDIS has higher power than all other algorithms when the nulls are conservative (i.e. $\mu_N < 0$), and ADDIS matches the power of SAFFRON without conservative nulls (i.e. $\mu_N = 0 $).

\section{Generalization of the discarding rule}\label{sec:general}
As we discussed before in \secref{addis}, one way to interpret what ADDIS is doing is that it is ``discarding'' the large $p$-values. We say ADDIS may be regarded as applying the ``discarding" rule to SAFFRON. Naturally, we would like to see whether the general advantage of this simple rule can be applied to other FDR control methods, and under more complex settings. We present the following generalizations and leave the details (formal setup, proofs) to supplement for interested readers. 

\begin{itemize}
	\item \textbf{Extension 1: non-adaptive methods with discarding }\\
	We derive the discarding version of LORD++ , which we would refer as D-LORD, in \secref{dislord}, with proved FDR control. 
	
	\item \textbf{Extension 2: discarding with asynchronous $p$-values}\\
	In a recent preprint, \citet{zrnic2019asynchronous} show how to generalize existing online FDR control methods to what they call the asynchronous multiple testing setting. They consider a doubly-sequential setup, where one is running a sequence of sequential experiments, many of which could be running in parallel, starting and ending at different times arbitrarily. In \secref{async}, we show how to unite the discarding rule from this paper with the ``principle of pessimism'' of \citet{zrnic2019asynchronous} to derive even more powerful asynchronous online FDR algorithms, which we would refer as ADDIS$_{\textnormal{async}}$.

	\item \textbf{Extension 3: Offline FDR control with discarding}\\
	In \secref{distorey}, we provide a new offline FDR control method called D-StBH, to show how to incorporate the discarding rule with the Storey-BH method, which is a common offline adaptive testing procedure \cite{Storey04,ramdas2019unified}. Note that in the offline setting, the discarding rule is fundamentally the same as the idea of \cite{zhao2018multiple}, which was only applied to non-adaptive global multiple testing. 
	
\end{itemize}

\begin{figure}[H]
	\centering
	\begin{subfigure}[t]{0.33\textwidth}
		\centering
		\includegraphics[width=0.9\linewidth]{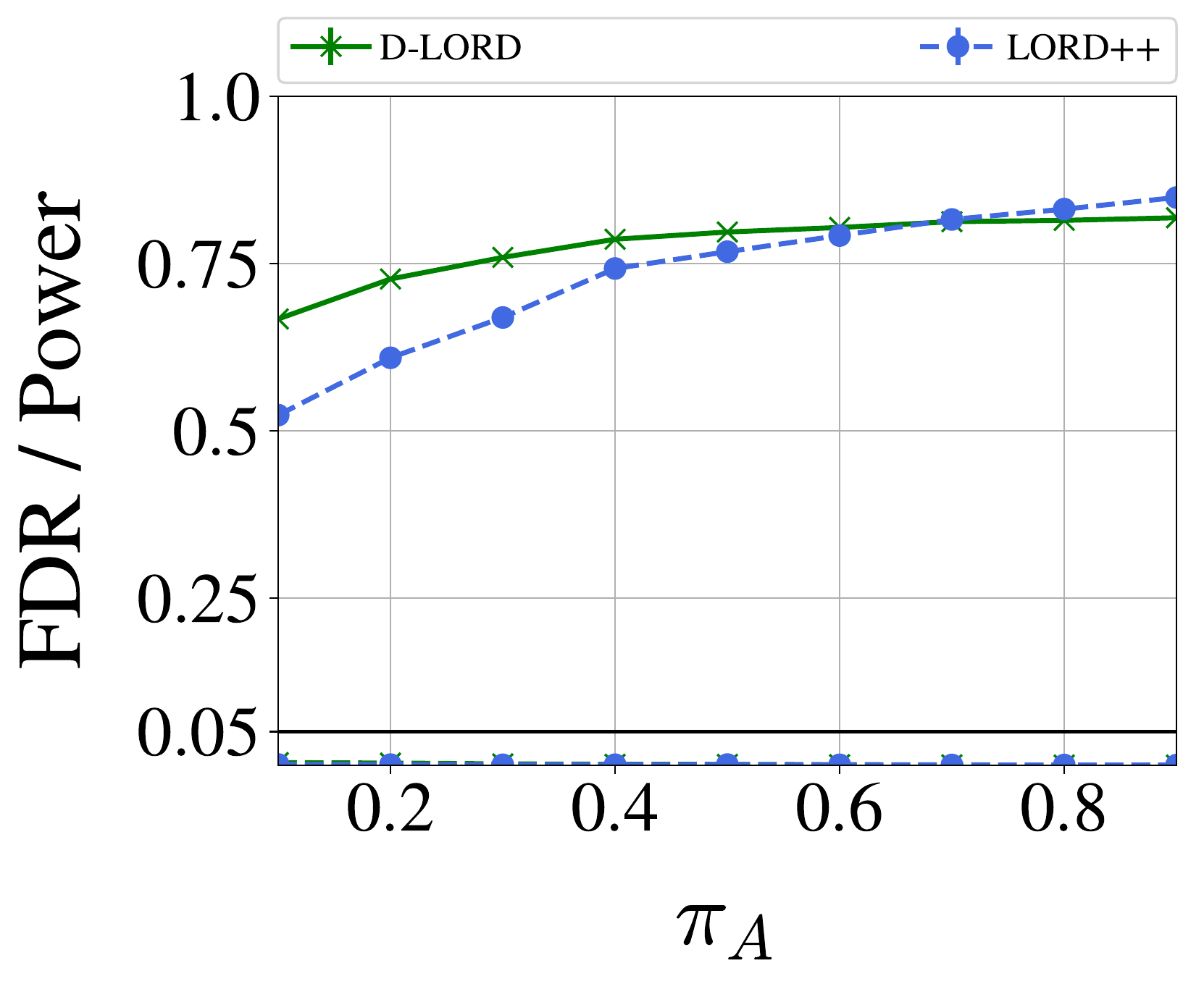}
		\caption{Extension 1}
	\end{subfigure}%
	\begin{subfigure}[t]{0.33\textwidth}
		\centering
		\includegraphics[width=0.9\linewidth]{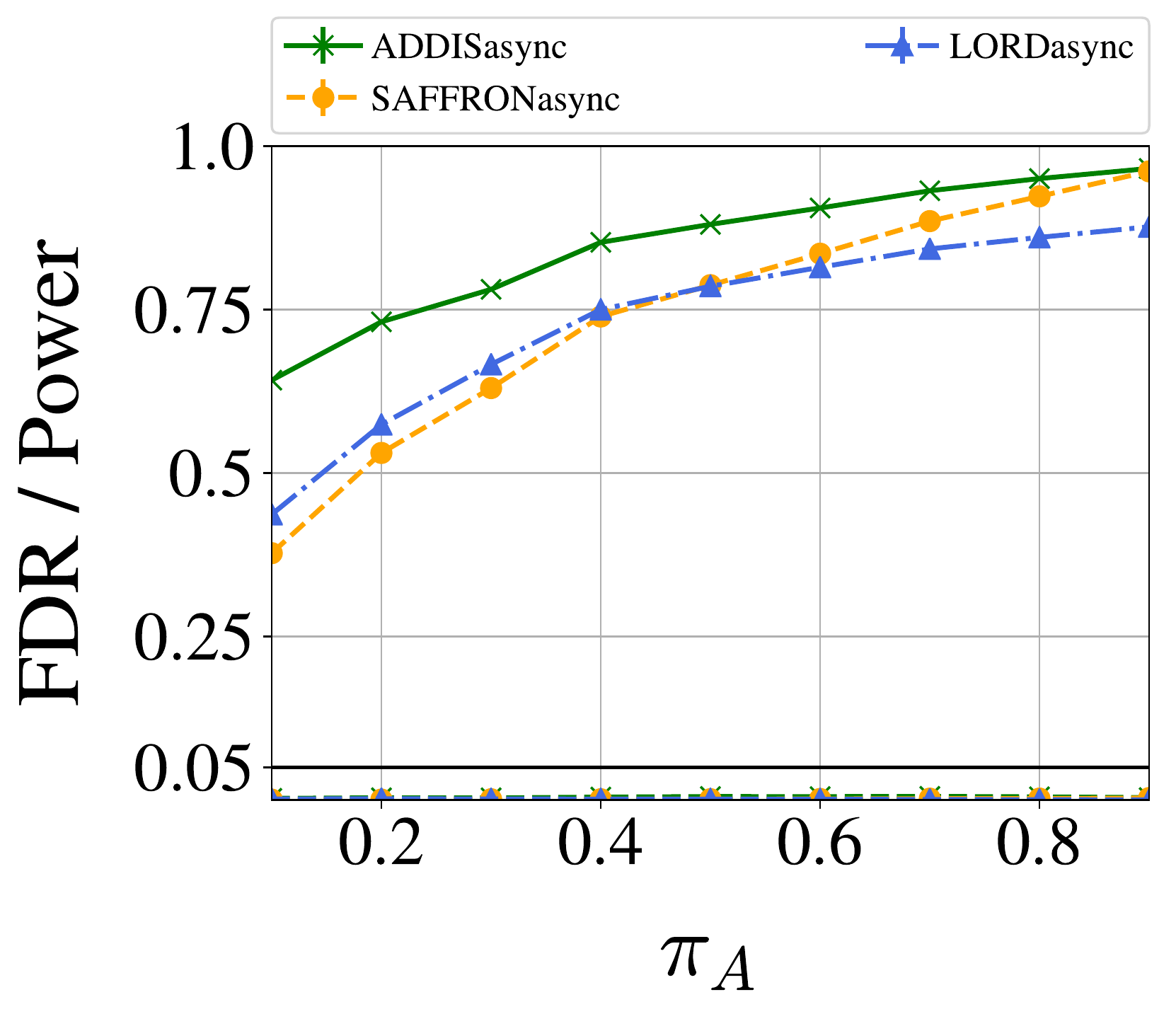}
		\caption{Extension 2}
	\end{subfigure}%
		\begin{subfigure}[t]{0.33\textwidth}
		\centering
		\includegraphics[width=0.9\linewidth]{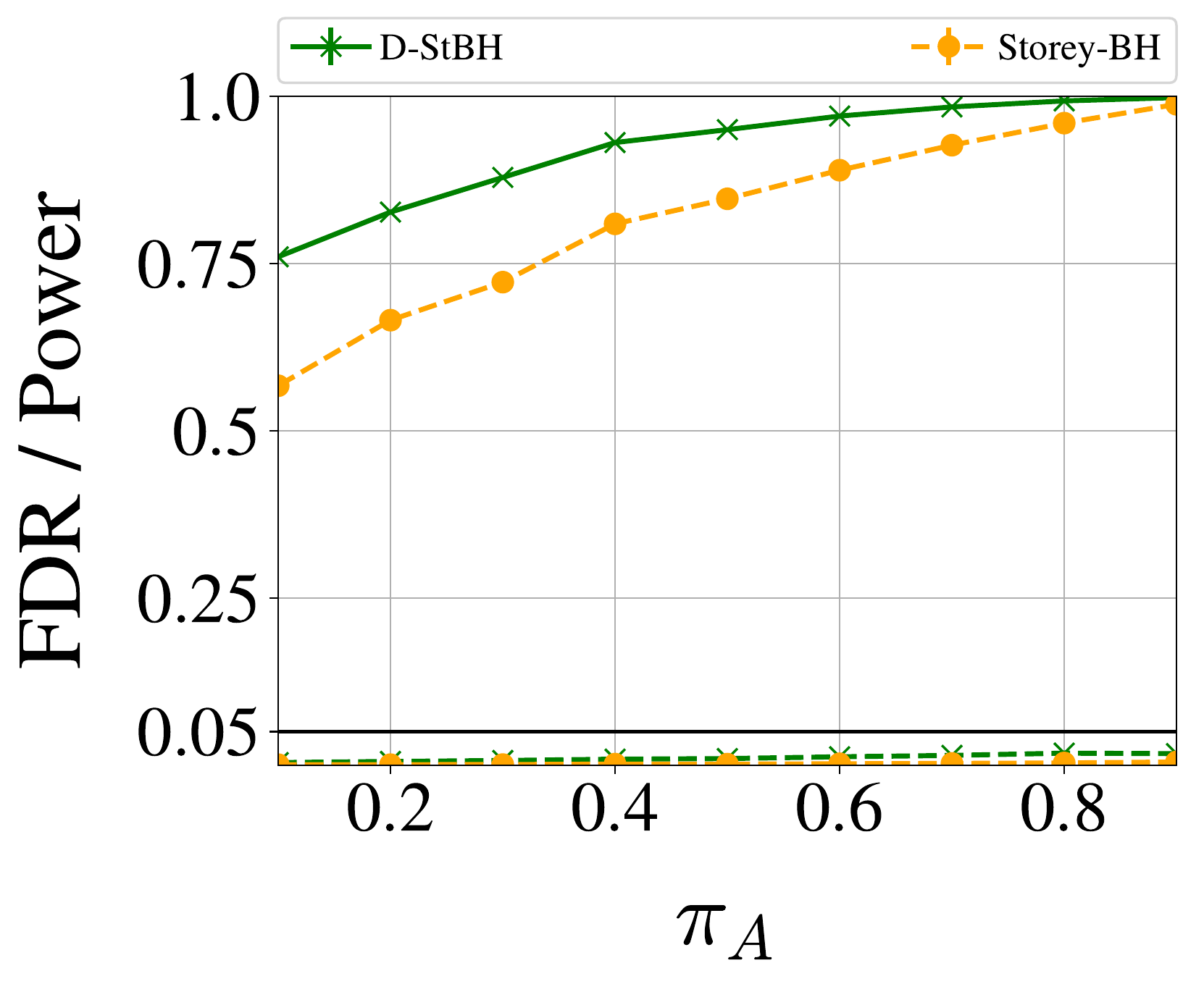}
		\caption{Extension 3}
	\end{subfigure}\\
	
	\begin{subfigure}[t]{0.33\textwidth}
		\centering
		\includegraphics[width=0.9\linewidth]{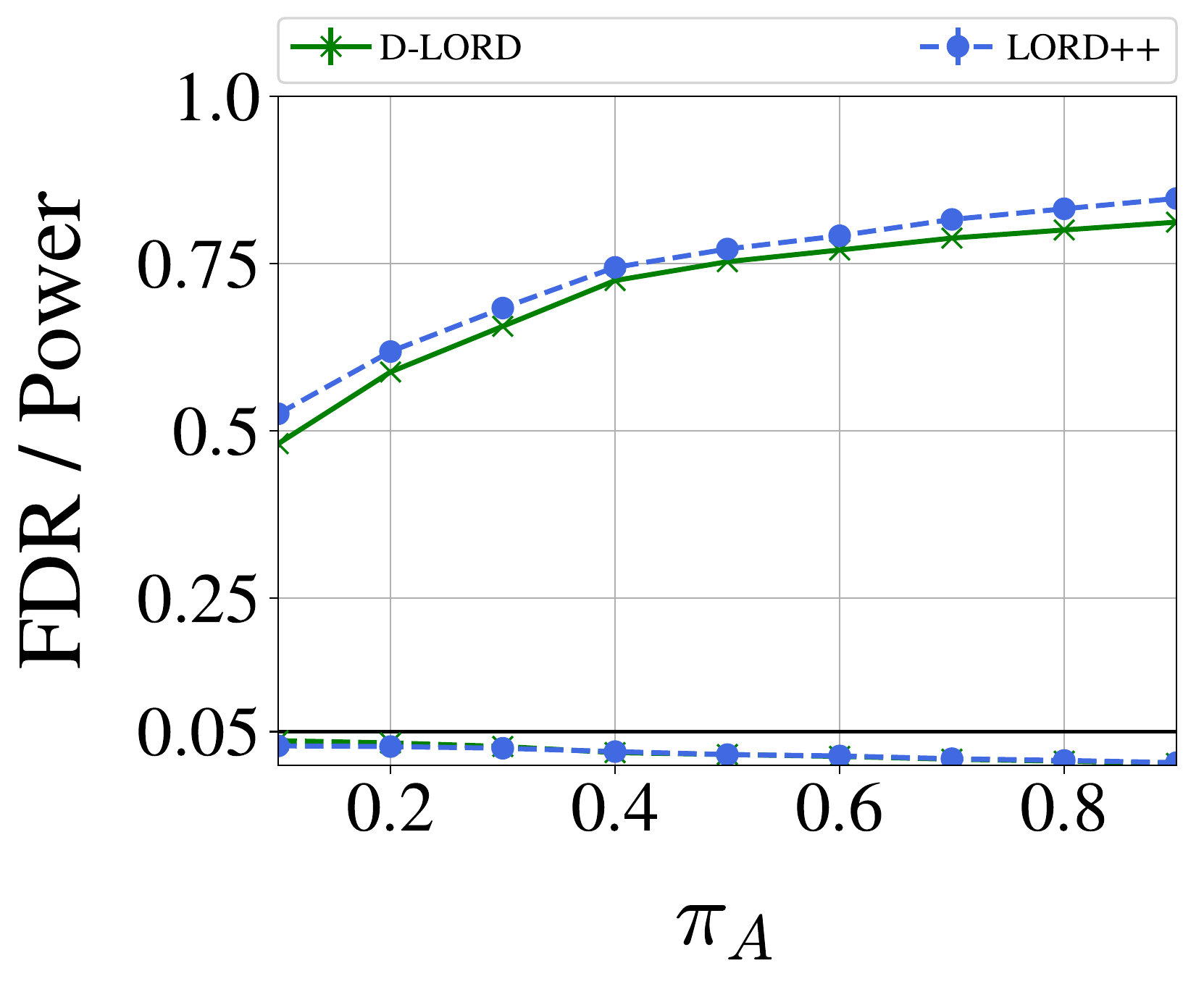}
		\caption{Extension 1}
	\end{subfigure}%
	\begin{subfigure}[t]{0.33\textwidth}
		\centering
		\includegraphics[width=0.9\linewidth]{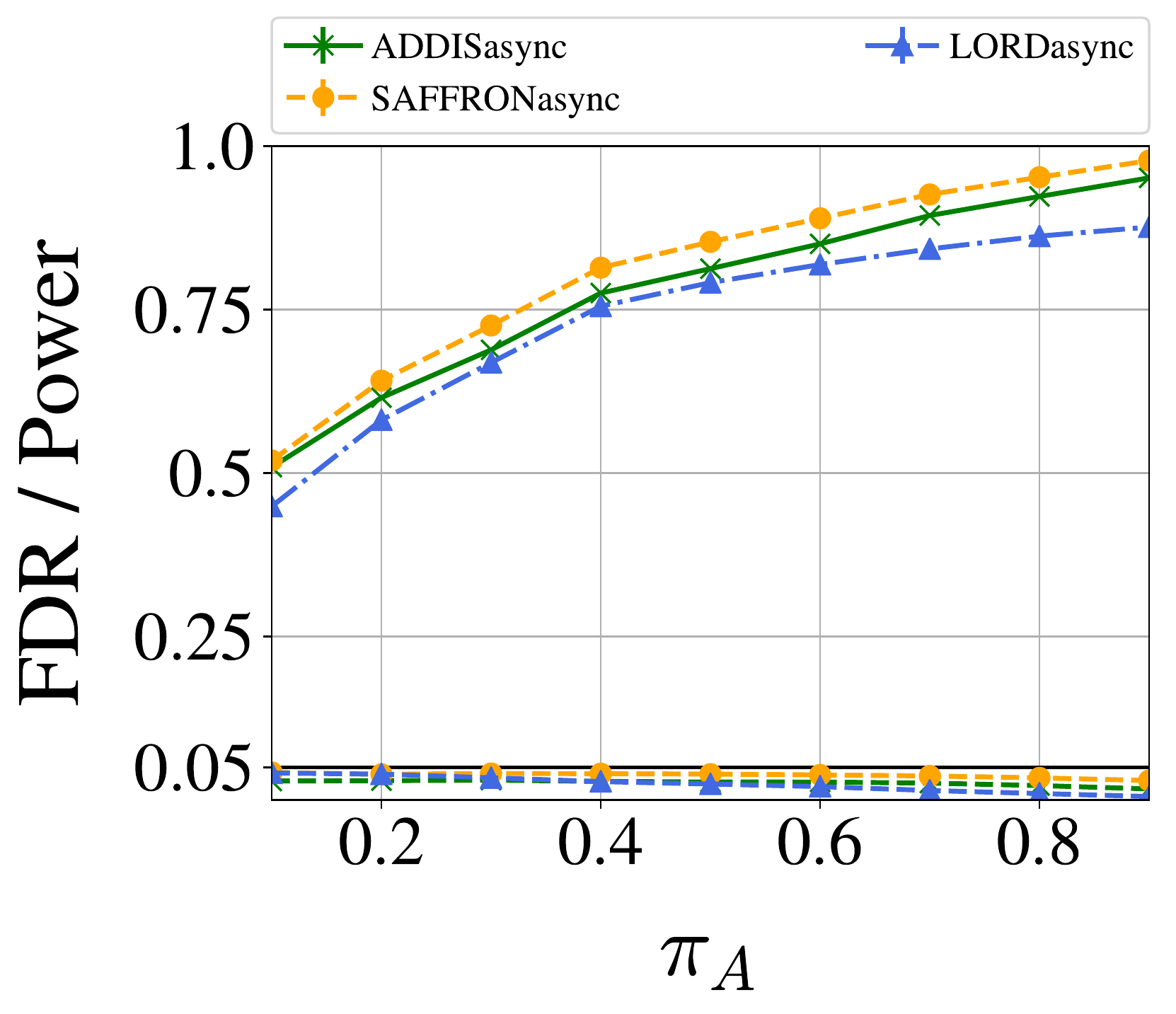}
		\caption{Extension 2}
	\end{subfigure}%
		\begin{subfigure}[t]{0.33\textwidth}
		\centering
		\includegraphics[width=0.9\linewidth]{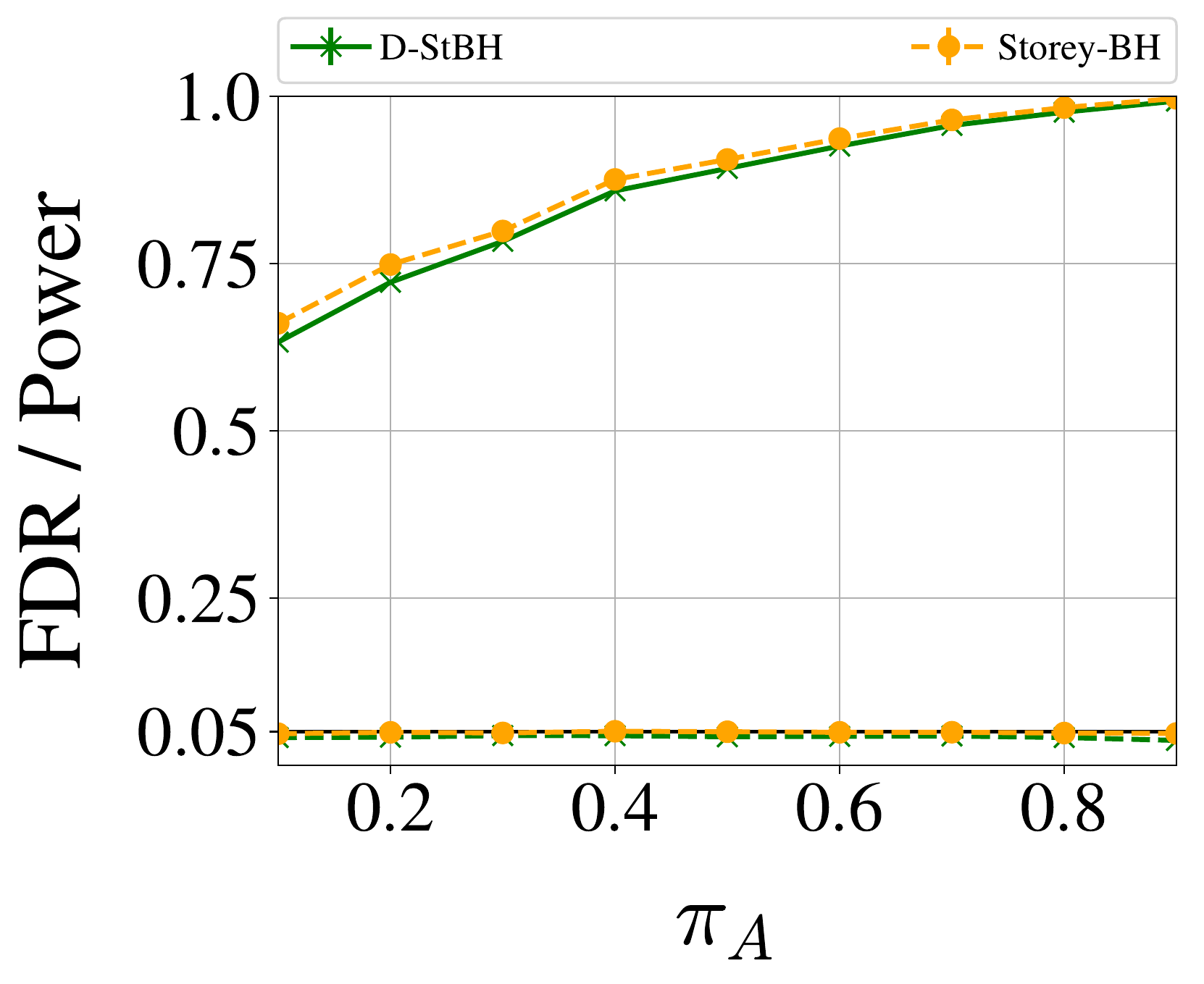}
		\caption{Extension 3}
	\end{subfigure}
	\caption{Statistical power and FDR versus fraction of non-null hypotheses $\pi_A$ for extended methods mentioned above at target FDR level $\alpha$ = 0.05 (solid black line). The $p$-values are drawn using the Gaussian model as described in the text, while we set $\mu_A = 3$ for all the figures, but $\mu_N = -1$ in plots (a, b, c), $\mu_N = 0$ in plots (d, e, f). We additionally set the finish time for the $j$-th test as $E_j \sim j-1 + \text{Geom}(0.5)$ in plots (b, e), which means the duration time of each individual tests independently follows Geometric distribution with succeed probability 0.5. }\label{fig:extend}
\end{figure}

The following simulation results in \figref{extend}, which are plotted in the same format as in \secref{simu}, show that those discarding variants (marked with green color) enjoys the same type of advantage over their non-discarding counterparts: they are consistently more powerful under settings with many conservative nulls and do not lose much power under settings without conservative nulls.

\section{Conclusion}\label{sec:conclusion}
In this work, we propose a new online FDR control method, ADDIS, to compensate for the unnecessary power loss of current online FDR control methods due to conservative nulls.  Numerical studies show that ADDIS is significantly more powerful than current state of arts, under settings with many conservative nulls, and rarely lose power under settings without conservative nulls. We also discuss the trade-off between adaptivity and discarding in ADDIS, together with some good heuristic of  how to balance them to obtain higher power.  In the end, we generalize the main idea of ADDIS to a simple but powerful rule  ``discarding'', and incorporate the rule with many current online FDR control methods under various settings to generate corresponding more powerful variants. For now, we mainly examine the power advantage of ADDIS algorithm with constant $\lambda$ and $\tau$, though for future work, how to choose time varying $\{\lambda_j\}_{j=1}^{\infty}$ and $\{\tau_j\}_{j=1}^{\infty}$ in a data adaptive matter with provable power increase is worthy of more attention. 

\subsection*{Acknowledgments}

We thank David Robertson for a careful reading and finding some typos in an early preprint.

\bibliographystyle{unsrtnat}
\bibliography{FDR_ADDIS}

\renewcommand{\thesection}{\Alph{section}.\arabic{section}}
\setcounter{section}{0}

\begin{appendices}
	
	\section{D-LORD: LORD++ with discarding}\label{sec:dislord}
Instead of applying discarding rule to LORD described in \cite{javanmard2017on}, we apply the discarding rule to its equivalent form LORD++ under the framework of GAI++ \cite{ramdas2017on} for theoretical simplicity, and call the resulted variant as D-LORD. Now consider uniformly conservative $p$-values as defined in \eqref{condition-conserve-def}, where the filtration $\Ft=  \sigma(R_{1:t-1}, S_{1:t-1})$. As before, we derive D-LORD from an empirical estimate of $\fdp^{*}$ defined in \eqref{fdp*}. Specifically, let 
\begin{equation}
\widehat{\FDP}_{\textnormal{D-LORD}}(t) \defn \frac{\sum_{j \leq t}\frac{\alpha_j}{\tau_j}\ident\{ P_j \leq \tau_j\}}{|R(t)|\vee 1}.
\end{equation}
Compare $\widehat{\FDP}_{\textnormal{D-LORD}}$ with the original estimator that LORD++ based upon
\begin{equation}
\widehat{\FDP}_{\textnormal{LORD++}}(t) \defn \frac{\sum_{j \leq t}\alpha_j }{|R(t)| \vee 1},
\end{equation}
we say $\widehat{\FDP}_{\textnormal{D-LORD}}$ is a better estimator, since with many  conservative null $p$-values, its numerator will be a much tighter estimate of $\sum_{j\leq t, j\in \nulls} \alpha_j$, compared with the naive estimate of LORD++ that is $\sum_{j\leq t} \alpha_j$. To see why this is true, just notice that the expectation of $\frac{\ident\{P_j \leq \tau_j\}}{\tau_j}$ will be much smaller than 1 for conservative null $p$-values.  We call an online FDR
algorithm as an instance of the ``D-LORD algorithm" if it updates $\alpha_t$ in a way such that it maintains the invariant $	\widehat{\FDP}_{\textnormal{D-LORD}}(t) \leq \alpha$ for all $t$.
We show how to ensure this invariant in a fully online fashion by providing an explicit instance of D-LORD with constant $\tau$ as the following D-LORD$^{*}$ algorithm. The simulation results in \secref{general} demonstrate the power advantage of D-LORD$^{*}$ over LORD++. 

\begin{algorithm}[H]
	\KwIn{FDR level $\alpha$, discarding threshold $\tau \in (0,1]$, sequence $\{\gamma_j \}_{j=0}^{\infty}$ which is nonnegative, nonincreasing and sums to one, initial wealth $W_0 \leq \alpha$.}
	\For{t=1, 2, \dots}{
		Reject the $t$-th null if $P_t \leq \alpha_t$, where $\alpha_t \defn \min\{\tau, \widetilde{\alpha}_t\},$ and \\ 
		$ \widetilde{\alpha}_{t}  \defn \tau\left(W_0\gamma_{S^t}  +(\alpha-W_0)\gamma_{S^t-\kappa_1^{*}}  + \alpha\sum_{j\geq 2} \gamma_{S^t - \kappa_j^{*}}\right).$\\
		
		Here, \\
		\ \ \ $\kappa_j = \min \{i\in [t-1]: \sum_{k\leq i}\ident\{P_k \leq \alpha_k\}\geq j\}$,
		\ \ \ $\kappa_j^{*} = \sum_{i\leq \kappa_j}\ident\{P_i \leq \tau\}$, 
		\ \ \ $S^{t} = \sum_{i <  t} \ident\{P_i \leq \tau\}$.
	}
	\caption{The D-LORD$^{*}$ algorithm}\label{dislord}
\end{algorithm}

Here we present the following theorem for error control of D-LORD. Recall the definition of uniformly conservative $p$-values  \eqref{condition-conserve-def}; and here we call a function $f_t(R_{1:t-1}, S_{1:t-1}): \{0,1\}^{2(t-1)} \to [0,1]$ as the ``monotonic'' function of the past if it is coordinatewise nondecreasing with regard $R_j$, and coordinatewise nonincreasing with regard $S_j$.
\begin{theorem}\label{thm:thmdislord}
	If the null $p$-values are uniformly conservative, and suppose we choose $ \tau_j \geq \alpha_j$ for each $j \in \mathbb{N}$, where $\alpha_j$ is the testing level for $j$-th hypothesis, then we have:\\
	\-\hspace{0.5cm}(a) any algorithm with  $\fdphat_{\textnormal{D-LORD}}(t) \leq \alpha$ for all $t
	\in \N$ also enjoys $\mfdr(t) \leq \alpha$ for all $t \in \N$.\\
	Further, if the null $p$-values are independent of each other and of
	the non-nulls, and for all $t$, $\alpha_t$ and $1-\tau_t$ are both monotonic functions of the past, then we additionally have:\\
	\-\hspace{0.5cm} (b) any algorithm with  $\fdphat_{\textnormal{D-LORD}}(t) \leq \alpha$ for all $t
	\in \N$ also enjoys $\fdr(t) \leq \alpha$ for all $t \in \N$.\\
	As an immediate corollary, D-LORD$^{*}$ (Algorithm \ref{dislord}) enjoys both mFDR and FDR control.
\end{theorem}
\noindent
The proof of \thmref{thmdislord} is presented in \secref{pfthmdislord}.

\section{D-StBH:  Storey-BH with discarding}\label{sec:distorey}
The discarding rule can also be applied to offline settings. Here we present the D-StBH, i.e. the discarding version of an adaptive offline FDR control method --- Storey-BH \cite{Storey04, ramdas2019unified}. Just as SAFFRON is an online analog of Storey-BH, ADDIS may be regarded as an online analog of D-StBH. 

Now we present the specific approach. Denote the number of hypotheses as $n$. Given targeted FDR level $\alpha$, user defined constants $\lambda<\tau \in (0,1]$, we define
\begin{align}\label{disstbh}
\fdphat_{\textnormal{D-StBH}}(s) \defn \frac{n \cdot s \cdot \widehat{\pi_0}}{(\sum_{j}\ident\{P_j \leq s \})\vee 1},
\ \textnormal{ where }\ \widehat{\pi_0} \defn \frac{1 + \sum_{i=1}^n \ident\{\lambda < P_i \leq \tau\} }{n(\tau-\lambda)}.
\end{align}
D-StBH then calculates $\widehat{s} \defn \max\{s: s\leq \tau , \fdphat_{\textnormal{D-StBH}}(s) \leq \alpha \}$,  and reject the set $\{i: P_i \leq \widehat{s}\}$. With many conservative nulls, we claim D-StBH would be more powerful than Storey-BH, since $\widehat{\pi}_0$ serves as a tighter estimator for the true $\pi_0 \defn |\nulls|/n$ in terms of expectation. As always, we present the error control of the new method under some reasonable assumptions, and the simulations demonstrating its power advantage in \secref{general}.

\begin{theorem}\label{thm:thmdisstorey}
	If $p$-values are independent with each other and the nulls are uniformly conservative as defined in \eqref{conserve-def}, then \textnormal{D-StBH} controls \fdr$\ $ at level $\alpha$.  
\end{theorem}	
\noindent \thmref{thmdisstorey} is proved in \secref{pfthmdisstorey} .

\section{Asynchronous setting}\label{sec:async}
Here we formalize the asynchronous setting. An asynchronous testing process consists of tests that start and finish at random times. Without loss of generality, one can take the starting times of each tests as $1, 2,\dots$, and refer them as $H_1, H_2, \dots$, and take the finish time of each tests as $E_1, E_2, \dots $ accordingly (let $E_t = j$, if $j \leq E_t < j+1$). Notice that $E_t$ may be bigger than $t$. One has to decide the testing level for $H_t$ at its starting time, with only information of tests that finished before time $t$.  It is worth mentioning that this framework is a generalization of the classical online FDR setting, since it reduces to the classical setting when $E_t = t$ for all $t$. We refer readers to \cite{zrnic2019asynchronous} for more detailed definition and discussion.

In the following of the section, we present the modified ADDIS algorithm under asynchronous setting, which we will refer as $\textnormal{ADDIS}_\textnormal{async}$. We derive the new method respectively from the following two empirical estimators for the oracle metric $\fdp^{*}$ for true $\fdp$, which is 
\begin{equation}
\widehat{\FDP}_{\textnormal{ADDIS}_\textnormal{async}}(t) \defn \frac{\sum_{j \leq t}\frac{\alpha_j}{(\tau_j-\lambda_j)}\left(\ident\{\lambda_j < P_j \leq \tau_j, E_j < t\}+\ident\{E_j\geq t\}\right)}{(\sum_{j\leq t}\ident\{P_j \leq \alpha_j, E_j < t\})\vee 1}.
\end{equation}

\noindent
As before, $\{\tau_j\}_{j=1}^{\infty}, \{\lambda_j\}_{j=1}^{\infty}, \{\alpha_j\}_{j=1}^{\infty}$ are some user defined sequences, where each terms is in range $[0,1]$. We use $P_t$ to refer the $p$-value that results from the test started at time $t$, which is not known at time $t$, but only at time $E_t$ (unless they are identical). Similarly,  $S_t, C_t, R_t$ are defined in the same way as \secref{addis}, to indicate whether the hypothesis started at time $t$ is selected, candidate of rejection, or rejected, respectively. Like $P_t $, they are also not known before time $E_t$. Additionally, denote $\R_t = \{i: E_i = t, R_i =1 \}$, $\C_t= \{i: E_i = t, C_i =1 \}$ and $\S_t= \{i: E_i = t, S_i =1\}$.  Correspondingly, denote $\R_{1:t} = \{\R_1, \dots, \R_t\}$, $\C_{1:t} = \{\C_1, \dots, \C_t\}$ and $\S_{1:t} = \{\S_1, \dots, \S_t\}$. As always, we refer the online FDR algorithm as ADDIS$_\textnormal{async}$ if it updates $\alpha_t$ to maintain the invariant $\widehat{\FDP}_{\textnormal{ADDIS}_\textnormal{async}}(t)\leq \alpha$ for all $t \in \N$.

Now we present explicit instance for  $\textnormal{ADDIS}_\textnormal{async}$ algorithm for fixed $\tau$ and $\lambda$.

\begin{algorithm}[H]
	\KwIn{FDR level $\alpha$, discarding threshold $\tau \in (0,1]$, candidate threshold $\lambda \in [0,\tau)$, sequence $\{\gamma_j \}_{j=1}^{\infty}$ which is nonnegative, nonincreasing and sums to one, initial wealth $W_0 \leq \alpha$.}
	\For{ $t =1, 2, \dots$}{
		Start $t$-th test with level $\alpha_t \defn \min\{\lambda, \widetilde{\alpha}_t\},$\\
		where $\widetilde{\alpha}_{t}  \defn (\tau-\lambda)\left(W_0\gamma_{S^t - C_{0}^{+}} +(\alpha-W_0)\gamma_{S^t-\kappa_1^{*}-C_{1}^{+}} + \alpha\sum_{j\geq 2} \gamma_{S^t - \kappa_j^{*}-C_{j}^{+}}\right)$.\\
		Here, $S^{t} = \sum_{i < t} (\ident\{P_i \leq \tau, E_i < t\} + \ident\{E_i \geq t\}),$\\
		$\ \ \ \ \ \ \ \ \  C_{j}^{+} = \sum_{i < t} \ident\{P_i \leq \lambda,\  \kappa_j+1\leq E_i < t\},$\\
		$\ \ \ \ \ \ \ \ \  \kappa_j = \min \{i\in [t-1]: \sum_{k \leq  t}  \ident\{P_k \leq \alpha_k, E_k\leq i\}\geq j\},$\\
		$\ \ \ \ \ \ \ \ \  \kappa_j^{*} = \sum_{i < t} \ident \{P_i \leq \tau,E_i \leq \kappa_j\}.$\\
	}
	\caption{The $\textnormal{ADDIS}^{*}_\textnormal{async}$ algorithm}\label{algoaddisasync}
\end{algorithm}

As always, we present the error control for the $\textnormal{ADDIS}_\textnormal{async}$, by proving theorem as the following. Firstly, we clarify the following terms.

Here, we say $P_i$ is uniformly conservative, if it satisfy the uniformly conservative condition defined in \eqref{condition-conserve-def}, with specified filtration $\F^{E_t - 1}$, where $\Ft = \sigma\{\R_{1:t-1}, \C_{1:t-1}, \S_{1:t-1}\}$. We insist that the thresholds $\tau_j, \lambda_j$ and $\alpha_j$ in $\textnormal{ADDIS}_\textnormal{async}$ are mappings from $(\R_{1:j-1}, \C_{1:j-1}, \S_{1:j-1})$ to $[0,1]$ for each $j \in \N$. Here, we say $f_t$ is a monotonic function of the past, if it is nondecreasing in $|\R_j|$ and $|\C_j|$, while nonincreasing in $|\S_j|$.

\begin{theorem}\label{thm:asyncaddis}
	If the null $p$-values are uniformly conservative, suppose we choose $ \tau_j > \lambda_j \geq \alpha_j$ for each $j \in \mathbb{N}$. Then we have:\\
	\-\hspace{0.5cm} (a) any algorithm with $\fdphat_{\textnormal{ADDIS}_\textnormal{async}}(t) \leq \alpha$ for all $t \in \N$ enjoys $\mfdr(t) \leq \alpha$ for all $t \in \N$. \\
	Next assume that the null $p$-values are independent of each other and of the non-nulls, and each $p$-value $P_t$ is independent of its decision time given $\F^{E_t-1}$. If $\alpha_t$, $\lambda_t$, $1-\tau_t$ are all designed to be monotonic functions of the past for all $t \in \N$, then we additionally have:\\ 
	\-\hspace{0.5cm} (b) any algorithm with $\fdphat_{\textnormal{ADDIS}_\textnormal{async}}(t) \leq \alpha$ for all $t \in \N$  enjoys $\fdr(t) \leq \alpha$ for all $t \in \N$.\\
	As an immediate corollary, ADDIS$^{*}_\textnormal{async}$ (Algorithm~\ref{algoaddisasync}) have both mFDR and FDR control.
\end{theorem}

\noindent
\thmref{asyncaddis} is proved using \lemref{asynclem}, which is a modified version of \lemref{addislem} in  \secref{addis}. The proof is presented in \secref{pfasync}.

\section{Proof of \lemref{estimates}}\label{sec:pfestimates}

Let $F_j$ denote the CDF of null $p$-value $P_j$, for fixed $b \in (0,1)$, let $h_j(a) = b F_j(a) - F_j(ab)$. Since $F_j$ is differentiable, let $f_j$ denote its density function, and notice that $f_j$ is monotonically increasing by the fact that $F_j$ is convex. Then we have that the derivative of $h_j$ is
\[
h_j^{\prime} (a) = \tau_j f_j(a) - b f_j(ab) \geq 0.
\]
Therefore, $h_j$ is increasing with $a$, which implies $h_j(a) \leq h_j(1)$. With simple rearrangement, we have
\[
\frac{\PP{ab < P_j \leq b}}{b(1-a)} \leq \frac{\PP{ P_j > a }}{(1-a)}
\]
as claimed.\\

\section{Proof of \thmref{addisthm}}\label{sec:pfaddisthm}
Part (a) of \thmref{addisthm} is proved using the the law of iterated expectations and the property of uniformly conservative null $p$-values as stated in \eqref{condition-conserve-def}. Specifically, taking iterated expectation by conditioning on $\{\Fj, S_j\}$ respectively for each $j \in \nulls$, we have

\begin{equation}\label{addiseq1}
\begin{split}
\EE{|\nulls\cap \R(t)|} & =  \sum_{j\leq t, j\in \nulls}\EE{\ident\{P_j \leq \alpha_j\}}\\
& = \sum_{j\leq t, j\in \nulls } \EE{\EEst{\ident\{P_j \leq \alpha_j\}}{S_j, \Fj }}\\
& \stackrel{(i)}{=}  \sum_{j\leq t, j\in \nulls } \EE{\EEst{\ident\{\frac{P_j}{\tau_j} \leq \frac{\alpha_j}{\tau_j}\}}{S_j = 1, \Fj}\PPst{S_j = 1}{\Fj}}\\
& =  \sum_{j\leq t, j\in \nulls } \EE{\EEst{\ident\{\frac{P_j}{\tau_j} \leq \frac{\alpha_j}{\tau_j}\}}{P_j\leq \tau_j, \Fj}\PPst{S_j = 1}{\Fj}},
\end{split}
\end{equation}	
where (i) is true since $\alpha_j < \tau_j$, therefore $P_j \leq \alpha_j$ implies $S_j =1$. Then, using the property of the uniformly conservative null $p$-values stated in \eqref{condition-conserve-def}, we have
\begin{equation}\label{addiseq2}
\begin{split}
&\sum_{j\leq t, j\in \nulls } \EE{\EEst{\ident\{\frac{P_j}{\tau_j} \leq \frac{\alpha_j}{\tau_j}\}}{P_j\leq \tau_j, \Fj}\PPst{S_j = 1}{\Fj}}\\
\leq &\sum_{j\leq t, j\in \nulls } \EE{ \frac{\alpha_j}{\tau_j}\PPst{S_j = 1}{\Fj}}\\
\leq &\sum_{j\leq t, j\in \nulls}\EE{\frac{\alpha_j}{\tau_j}\EEst{ \frac{\ident\{\theta_j\tau_j<P_j\}}{(1-\theta_j)}}{P_j\leq \tau_j, \Fj}\PPst{S_j = 1}{\Fj}},
\end{split}
\end{equation}
where $\theta_j \equiv \lambda_j / \tau_j$. Next, using the fact that $\alpha_j,\lambda_j$ and $ \tau_j$ are measurable with regard $\Fj$ for all $j \in \N$, the RHS of \eqref{addiseq2} equals 
\begin{equation}\label{addiseq3}
\begin{split}
& \sum_{j\leq t, j\in \nulls}\EE{\EEst{\frac{\alpha_j}{\tau_j}\frac{\ident\{\theta_j\tau_j<P_j\}}{(1-\theta_j)}}{P_j\leq \tau_j, \Fj}\PPst{S_j = 1}{\Fj}}\\
= &\sum_{j\leq t, j\in \nulls}\EE{\EEst{\alpha_j\frac{\ident\{\lambda_j<P_j\leq \tau_j\}}{(\tau_j-\lambda_j)}}{S_j = 1, \Fj}\PPst{S_j = 1}{\Fj}}\\
= &\sum_{j\leq t, j\in \nulls}\EE{\EEst{\alpha_j\frac{\ident\{\lambda_j<P_j\leq \tau_j\}}{(\tau_j-\lambda_j)}}{S_j, \Fj}}\\
\stackrel{(ii)}{=} & \sum_{j\leq t, j\in \nulls} \EE{\alpha_j\frac{\ident\{\lambda_j<P_j\leq \tau_j\}}{(\tau_j-\lambda_j)}} \stackrel{(iii)}{=} \EE{\sum_{j\leq t, j\in \nulls}\alpha_j\frac{\ident\{\lambda_j<P_j\leq \tau_j\}}{(\tau_j-\lambda_j)}},
\end{split}
\end{equation}
\noindent
where (ii) is again obtained using law of the iterated expectations; and (iii) is obtained using the linearity of expectation. Therefore, combine the results above, we have 
\begin{equation}\label{(a)}
\EE{|\nulls\cap R(t)|} \leq\EE{\sum_{j \leq t, j\in \nulls}\alpha_j\frac{\ident\{\lambda_j < P_j \leq \tau_j\}}{(\tau_j-\lambda_j)}}.
\end{equation}
Furthermore, since
\[
\fdphat_{\textnormal{\textnormal{ADDIS}}}(t)\leq\alpha \Rightarrow \sum_{j\leq t} \alpha_j \frac{\ident\{\lambda_j<P_j\leq\tau_j\}}{(\tau_j-\lambda_j)} \leq \alpha (|R(t)|\vee 1), 
\]
take expectation on each side and use \eqref{(a)}, we have $\mfdr(t)\leq \alpha$ as claimed.\\

\noindent
Next, in order to prove part (b), we need \lemref{addislem} in the following, which is a modified version of ``reverse super-uniformity lemma'' in \cite{ramdas2018saffron}. Recall the definition of ``monotonic (neg-montonic) function of the past" in \ref{mono}, we present \lemref{addislem} as follows.

\begin{lemma}\label{lem:addislem}
	Assume that the $p$-values $P_1,P_2,\dots$ are independent and let $g: \{0,1\}^T \to \mathbb{R}$ be any coordinatewise nondecreasing function, and assume $\alpha_t, \lambda_t$ and $1- \tau_t$ are all monotonic function of the past as defined in \ref{mono}, while satisfying the constraints $\alpha_t \leq \lambda_t < \tau_t$ for all $t$. Then, for any index $t \leq T$ such that $t \in \nulls$, we have:
	\begin{align*}
	\EEst{\frac{\alpha_t\ident\{\lambda_t<P_t\leq\tau_t\}}{(\tau_t-\lambda_t)(g(R_{1:T})\vee 1)}}{\Ft, S_t=1} &\geq \EEst{\frac{\alpha_t}{\tau_t (g(R_{1:T})\vee 1)}}{\Ft, S_t=1}
	\\&\geq\EEst{\frac{\ident\{P_t\leq \alpha_t\}}{g(R_{1:T})\vee 1}}{\Ft, S_t = 1}.
	\end{align*}
\end{lemma}
\noindent
The proof of  \lemref{addislem} is deferred in \secref{pfaddislem}. 

Now, taking iterated expectations similarly as in the proof of part (a), we  obtain the following:
\begin{align}\label{addiseq4}
\fdr(t) = \EE{\fdp(t)} &= \EE{\frac{|\nulls\cap R(t)|}{|R(t)|\vee 1}}=\sum_{j\leq t, j\in \nulls}\EE{\frac{\ident\{P_j\leq\alpha_j\}}{|R(t)|\vee 1}}\nonumber\\
&=  \sum_{j\leq t, j\in \nulls}\EE{\EEst{\frac{\ident\{P_j\leq\alpha_j\}}{|R(t)|\vee 1}}{S_j, \Fj}} \nonumber\\
&=  \sum_{j\leq t, j\in \nulls}\EE{\EEst{\frac{\ident\{P_j\leq\alpha_j\}}{|R(t)|\vee 1}}{S_j=1, \Fj}\PPst{S_j = 1}{\Fj}} 
\end{align}
Under the independence and monotonicity assumptions of part (b), and notice that $|R(t)| = \sum_{i=1}^{t} R_i$ is a coordinatewise nondecreasing function with regard $R_{1:t}$, we use \lemref{addislem} to obtain the following: 
\begin{align}\label{addiseq5}
&\sum_{j\leq t, j\in \nulls}\EE{\EEst{\frac{\ident\{P_j\leq\alpha_j\}}{|R(t)|\vee 1}}{S_j=1, \Fj}\PPst{S_j = 1}{\Fj}}\nonumber\\
\leq & \sum_{j\leq t, j\in \nulls}\EE{\EEst{\frac{\alpha_j}{\tau_j(|R(t)|\vee 1)}}{\S_j = 1, \Fj}\PPst{\S_j = 1}{\Fj}}
\nonumber\\
\leq & \sum_{j\leq t, j\in \nulls}\EE{ \EEst{\frac{\alpha_j\ident\{\lambda_j<P_j\leq\tau_j\}}{(\tau_j-\lambda_j)(|R(t)|\vee 1)}}{S_j=1, \Fj}\PPst{S_j=1}{\Fj}}
\end{align}
Again using the law of iterated expectation and the linearity of expectation, we have the RHS of \eqref{addiseq5} equals 
\begin{align}\label{addiseq6}
&\sum_{j\leq t, j\in \nulls}\EE{ \EEst{\frac{\alpha_j\ident\{\lambda_j<P_j\leq\tau_j\}}{(\tau_j-\lambda_j)(|R(t)|\vee 1)}}{S_j, \Fj}}\nonumber\\
= &\sum_{j\leq t, j\in \nulls}\EE{ \frac{\alpha_j\ident\{\lambda_j<P_j\leq\tau_j\}}{(\tau_j-\lambda_j)(|R(t)|\vee 1)}}\nonumber\\
= & \EE{\frac{1}{|R(t)|\vee 1}\sum_{j\leq t, j\in \nulls } \alpha_j \frac{\ident\{\lambda_j<P_j\leq\tau_j\}}{(\tau_j-\lambda_j)}},
\end{align}
which is no larger than $\EE{\fdphat_{\textnormal{ADDIS}}(t)}\leq \alpha$ by the definition of $\fdphat_{\textnormal{ADDIS}}(t)$. Therefore, combine \eqref{addiseq4}, \eqref{addiseq5} and \eqref{addiseq6}, we have $\fdr(t) \leq\alpha$ as claimed. 

Finally, we justify for the corollary that  $\textnormal{ADDIS}^{*}$ have mFDR and FDR control. Firstly, from Algorithm~\ref{addisalgo}, we know that $\textnormal{ADDIS}^{*}$  makes sure $\tau>\lambda\geq\alpha_j$ for all $j$, and constant $\lambda$  and $1-\tau$ is obviously monotonic  function of the past, while $\alpha_t$ being a monotonic function of the past for all $t$ is verified in \secref{verify}.  Then, from the definition of sequence $\{\gamma_j\}^{\infty}_{j=0}$, after simple rearrangement, we  have $\fdphat_{\textnormal{ADDIS}}(t) \leq \alpha$ holds true. Therefore, $\textnormal{ADDIS}^{*}$ satisfy all the requirements in the theorem, thus having error control under corresponding assumptions of $p$-values.

\subsection{Proof of  \lemref{addislem}}\label{sec:pfaddislem}
We use a technique of constructing a hallucinated vector, similar to \cite{ramdas2018saffron}, to prove  \lemref{addislem}. Specifically, to prove the first part of the inequality, first fix the time $t$, and then construct a hallucinated vector $\widetilde{P}$, such that for each $i \in \N$, 
\begin{equation}
\widetilde{P}_i = \tau_i \cdot \ident\{i=t\} + P_i \cdot \ident\{i\neq t\}.
\end{equation}

Denote the corresponding hallucinated testing levels, candidate levels and selected levels resulting from $\{\widetilde{P}_i\}$ as $\{\widetilde{\alpha}_i\}$, $\{\widetilde{\lambda}_i\}$ and  $\{\widetilde{\tau}_j\}$ respectively. Similarly, we define the corresponding hallucinated indicator variables as 
\[
\widetilde{S}_i = \ident\{\widetilde{P}_i\leq \widetilde{\tau}_i\},\ \ \  \widetilde{C}_i=\ident\{\widetilde{P}_i\leq{\widetilde{\lambda}_i}\},\ \ \ \ \widetilde{R}_i = \ident\{\widetilde{P}_i\leq \widetilde{\alpha}_i\}.
\]
Given $\lambda_t<P_t \leq \tau_t$, we have $\widetilde{S}_t = S_t = 1, \widetilde{R}_t = R_t = 0, \widetilde{C}_t = C_t = 0$. Therefore, $R_{1:T} = \widetilde{R}_{1:T}$, and particularly $\widetilde{R}_{1:T}$ is independent of $P_t$. These facts lead to:
\begin{equation*}
\begin{split}
&\ \ \ \ \ \  \EEst{\frac{\alpha_t\ident\{\lambda_t<P_t\leq \tau_t\}}{(\tau_t-\lambda_t)(g(R_{1:T})\vee 1)}}{S_t=1, \Ft}\\
& = \EEst{\frac{\alpha_t\ident\{\lambda_t<P_t\leq \tau_t\}}{(\tau_t-\lambda_t)(g(\widetilde{R}_{1:T})\vee 1)}}{S_t=1, \Ft}\\
&\stackrel{(i)}{=} \EEst{\frac{\alpha_t}{ \tau_t (g(\widetilde{R}_{1:T})\vee 1)}}{S_t = 1, \Ft} \EEst{\frac{\ident\{\lambda_t<P_t\leq \tau_t\}}{ (1-\lambda_t/\tau_t)}}{S_t = 1, \Ft} \\
&\stackrel{(ii)}{\geq} \EEst{\frac{\alpha_t}{\tau_t (g(\widetilde{R}_{1:T})\vee 1)}}{S_t = 1, \Ft},
\end{split}
\end{equation*}
where (i) is obtained from the fact that $\widetilde{R}_{1:T}$ is independent of $P_t$, and that  $\lambda_t, \tau_t, f_t$ are measurable with regard $\Ft$; (ii) is obtained using the property of uniformly conservative null $p$-values stated in \eqref{condition-conserve-def}. 

Under the construction of hallucinated variables, if $S_t = 1$, then $\widetilde{R}_i \leq R_i $ for all $i \in \N$. This statement follows by the monotonicity of $\{\alpha_i\}$ and $\{\lambda_i\}$, and the neg-monotonicity of $\{\tau_i\}$. Notice that for all $i<t$, we have $\ \widetilde{S}_i=S_i,\ \widetilde{R}_i = R_i,\ \widetilde{C}_i=C_i.$ Therefore, we may infer that $\alpha_i = \widetilde{\alpha}_i$, $\lambda_i = \widetilde{\lambda}_i$ and $\tau_i = \widetilde{\tau}_i$ for all $i \leq t$.
Since $\widetilde{S}_t = S_t = 1, \widetilde{R}_t = \widetilde{C}_t = 0$, that is $\widetilde{S}_t = S_t$, $\widetilde{C}_t \leq C_t$, $\widetilde{R}_t \leq R_t$. Therefore we have that $\widetilde{\alpha}_{t+1} \leq \alpha_{t+1}$, $\widetilde{\lambda}_{t+1} \leq \lambda_{t+1}$ and $\widetilde{\tau}_{t+1} \geq \tau_{t+1}$, which lead to   $\widetilde{R}_{t+1}\leq R_{t+1}$, $\widetilde{C}_{t+1}\leq C_{t+1}$ and $\widetilde{S}_{t+1}\geq S_{t+1}$ and so on. Recursively, we deduce $\widetilde{R}_{t+1}\leq R_{t+1}$  for all $i > t$. Since $g$ is a coordinatewise increasing function, we have
\begin{equation}
\EEst{\frac{\alpha_t}{\tau_t (g(\widetilde{R}_{1:T})\vee 1)}}{S_t = 1, \Ft} \leq \EEst{\frac{\alpha_t}{\tau_t (g(R_{1:T})\vee 1)}}{S_t = 1, \Ft}
\end{equation}
Hence, we proved the first part of inequality in \lemref{addislem}. \\

\noindent
To prove the second part of the inequality, alternatively, for all  $t \in \N$, we let $\widetilde{P}_i = P_i \cdot \ident\{i\neq t\}$, and define $\widetilde{\alpha}_i, \widetilde{\lambda}_i$, $\widetilde{\tau}_i$ and  $\widetilde{S}_i$, $\widetilde{C}_i$, $\widetilde{R}_i$ in same way as before. 

On the other hand, given $P_t \leq \alpha_t$, we have $\widetilde{S}_t = S_t = \widetilde{R}_t=R_t=\widetilde{C}_t= C_t = 1$. Therefore, $R_{1:T} = \widetilde{R}_{1:T}$, and particularly  $\widetilde{R}_{1:T}$ is independent of $P_t$. Again, we  have:
\begin{equation*}
\begin{split}
&\ \ \ \ \ \EEst{\frac{\ident\{P_t\leq \alpha_t\}}{g(R_{1:T})\vee 1}}{S_t = 1, \Ft}= \EEst{\frac{\ident\{P_t\leq \alpha_t\}}{g(\widetilde{R}_{1:T})\vee 1}}{S_t = 1, \Ft}\\
& \stackrel{(i)}{=} \EEst{\frac{\alpha_t}{\tau_t (g(\widetilde{R}_{1:T})\vee 1)}}{S_t = 1, \Ft} \EEst{\frac{\ident\{P_t\leq \alpha_t\}}{\alpha_t/\tau_t}}{S_t = 1, \Ft}\\
&\stackrel{(ii)}{\leq} \EEst{\frac{\alpha_t}{\tau_t (g(\widetilde{R}_{1:T})\vee 1)}}{S_t = 1, \Ft}
\stackrel{(iii)}{\leq} \EEst{\frac{\alpha_t}{\tau_t (g(R_{1:T})\vee 1)}}{S_t = 1, \Ft},
\end{split}
\end{equation*}
where (i) is obtained from the fact that $\widetilde{R}_{1:T}$ is independent of $P_t$, and that  $\lambda_t, \tau_t, f_t$ are measurable with regard $\Ft$; and (ii) is true due to the property of uniformly conservative $p$-values stated in  \eqref{condition-conserve-def} ; and finally, (iii) is true from the similar logic in the proof of first part.

These concludes the proof of the second part of inequality in \lemref{addislem}.

\subsection{Verify $\alpha_t$ in ADDIS$^{*}$ is a monotonic function of the past}\label{sec:verify}

In applying \thmref{addisthm} to prove that $\textnormal{ADDIS}^{*}$ controls the FDR, it is assumed that $\textnormal{ADDIS}^{*}$ is a monotonic rule, meaning that $\alpha_t$ is a monotonic function of the past as defined in \ref{mono}. Here we justify for this claim. In $\textnormal{ADDIS}^{*}$, we assume $\lambda$ and $\tau$ is constant, however the same arguments can be applied if they change at every step, but are predictable as stated in \secref{addis} of the main paper.

We will prove this argument by proving that $\alpha_t$ in $\textnormal{ADDIS}^{*}$ satisfy some equivalent argument of monotonicity defined in \ref{mono}. Consider some $(R_{1:t-1},C_{1:t-1}, S_{1:t-1})$ and $(\widetilde{R}_{1:t-1},\widetilde{C}_{1:t-1}, \widetilde{S}_{t-1})$ for a fixed $t$. We will accordingly denote all relevant variables in the $\textnormal{ADDIS}^{*}$ alogorithm which result in $(R_{1:t-1},C_{1:t-1}, S_{1:t-1})$ and $(\widetilde{R}_{1:t-1},\widetilde{C}_{1:t-1}, \widetilde{S}_{1:t-1})$, e.g. $\alpha_t$ and $\widetilde{\alpha}_t$, respectively. We say $(\widetilde{R}_{1:t-1},\widetilde{C}_{1:t-1}, \widetilde{S}_{1:t-1})\succeq (R_{1:t-1},C_{1:t-1}, S_{t-1})$ if and only if, for each $i\leq t-1$, one of the following holds:
\begin{enumerate}
	\item[(1)] $R_i = \widetilde{R}_i$, $C_i = \widetilde{C}_i$, and $R_i = \widetilde{R}_i$;
	\item[(2)] $R_i = 0$, $C_i = 0$, $S_i = 1$, and $\widetilde{R}_i = 0$, $\widetilde{C}_i = 1$, $\widetilde{S}_i=1;$
	\item[(3)] $R_i = 0$, $C_i = 0$, $S_i = 1$, and $\widetilde{R}_i = 1$, $\widetilde{C}_i = 1$, $\widetilde{S}_i = 1;$
	\item[(4)] $R_i = 0$, $C_i = 1$, $S_i = 1$, and $\widetilde{R}_i = 1$, $\widetilde{C}_i = 1$, $\widetilde{S}_i = 1.$
	\item[(5)] $R_i = 0$, $C_i = 0$, $S_i = 1$, and $\widetilde{R}_i = 0$, $\widetilde{C}_i = 0$, $\widetilde{S}_i = 0;$
\end{enumerate}

\noindent
Taking into account the possible relations between indicators for rejection, candidacy and tester, one may notice the fact that $S_i \geq C_i \geq R_i$ for each $i$. Then the monotonicity defined in \ref{mono} of a function $\alpha_t$ is equivalent to the statement that $(\widetilde{R}_{1:t-1},\widetilde{C}_{1:t-1}, \widetilde{S}_{1:t-1})\succeq (R_{1:t-1},C_{1:t-1}, S_{t-1})$ implies $\widetilde{\alpha}_t \geq \alpha_t$. Therefore, we will instead prove that this equivalent statement holds for $\alpha_t$ in $\textnormal{ADDIS}^{*}$ for each $t \in \N$. 
Specifically, recall the forms of $\alpha_t$ in $\textnormal{ADDIS}^{*}$:
\begin{equation}\label{alphat}
\begin{split}
&\alpha_t \defn \min\{ \lambda, \widehat{\alpha}_t\},\\ &\text{~ where ~}\widehat{\alpha}_t \defn (\tau-\lambda)\left(W_0\gamma_{S^t - C_{0+}} + (\alpha -
W_0) \gamma_{S^t-\kappa_1^{*} - C_{1+}} 
+  \alpha \sum_{j \geq 2}
\gamma_{S^t-\kappa_j^{*} - C_{j+}}\right).
\end{split}
\end{equation}
\noindent
We would like to prove that, given $(\widetilde{R}_{1:t-1},\widetilde{C}_{1:t-1},\widetilde{S}_{1:t-1}) \succeq (R_{1:t-1},C_{1:t-1},S_{1:t-1})$, we have $\widetilde{\alpha}_t \geq \alpha_t$. First, notice that in \eqref{alphat}, the index $S^t-\kappa_j^{*} - C_{j+}$ is the number of non-candidate testers (i.e. $\{i :S_i=1, C_i =0\}$) between the $j$-th rejection before time $t$ and time $t$. Provided with $(\widetilde{R}_{1:t-1},\widetilde{C}_{1:t-1},\widetilde{S}_{1:t-1}) \succeq (R_{1:t-1},C_{1:t-1},S_{1:t-1})$, we must have that $(R_{1:t-1},C_{1:t-1},S_{1:t-1})$ never contains less non-candidate testers or more rejections compared to  $(\widetilde{R}_{1:t-1},\widetilde{C}_{1:t-1},\widetilde{S}_{1:t-1})$, from the definition of $(\widetilde{R}_{1:t-1},\widetilde{C}_{1:t-1},\widetilde{S}_{1:t-1}) \succeq (R_{1:t-1},C_{1:t-1},S_{1:t-1})$ above. Additionally, notice that the sequence $\{\gamma_j\}_{j=0}^{\infty}$ is nonincreasing and nonnegative, and $W_0,\ \alpha -W_0$ and $\tau-\lambda$ in \eqref{alphat} are strictly positive by construction. Therefore, the sum of the terms $\gamma_{S^t-\kappa_j^{*} - C_{j+}}$ contributing to $\alpha_t$ is at most as great as the the sum of the terms $\gamma_{\widetilde{S}^t-\widetilde{\kappa}_j^{*} - \widetilde{C}_{j+}}$, and the same holds for the terms with $W_0$ and $(\alpha - W_0)$. Consequently, we have $\widetilde{\alpha}_t \geq \alpha_t$. Therefore, ADDIS$^{*}$ is a monotonic rule as claimed.

\section{Proof of \thmref{stopping}}\label{sec:pfstopping}

Using a similar technique to \cite{zrnic2019asynchronous}, we prove this theorem by constructing a process which behaves similarly to a submartingale, so that we could obtain a result by mimicking optimal stopping. Specifically, for $t \in \mathbb{N}$, define the process $A(t)$ as:

\begin{align*}
A(t) & \defn \sum_{i\leq t, i \in \nulls} \left(- \ident\{P_j \leq \alpha_j \} + \frac{\alpha_j}{(\tau_j-\lambda_j)} \ident\{\lambda_j \leq P_j < \tau_j\}  \right),
\end{align*}
where we take $A(0) = 0$. Denote $R(t)$ as the set of all rejections made by time $t$, and $V(t)$ as the set of false rejections made by time $t$. Then, we  bound

\begin{align*}
A(t) &= \sum_{i\leq t, i \in \nulls} \left(- \ident\{P_j \leq \alpha_j \} + \frac{\alpha_j}{(\tau_j-\lambda_j)} \ident\{\lambda_j \leq P_j < \tau_j\}  \right)\\
& \leq - |V(t)| + \sum_{j\leq t} \frac{\alpha_j}{(\tau_j-\lambda_j)} \ident\{\lambda_j < P_j \leq \tau_j\} \\
& =  \alpha (|R(t)| \vee 1)	 - V(t) + \sum_{j\leq t} \frac{\alpha_j}{(\tau_j-\lambda_j)} \ident\{\lambda_j < P_j \leq \tau_j\} - \alpha (|R(t)| \vee 1)\\	
&\stackrel{(i)}{\leq} \alpha (|R(t)|\vee 1)	 - V(t),\\
\end{align*}
where (i) is obtained using the fact that $\fdp_{\textnormal{ADDIS}}(t) \leq \alpha$ for all $t$. Therefore, if we can prove $A(T_\textnormal{stop}) \geq 0$ for any stopping time $T_\textnormal{stop}$ with finite expectation, then we instantly obtain $\alpha |R(T_\textnormal{stop})|\geq V(T_\textnormal{stop})$. Taking expectation on both side, and rearranging the terms, we  obtain  $\mfdr(T_\textnormal{stop}) \leq \alpha$ as claimed. 

In order to prove $A(T_\textnormal{stop}) \geq 0$ for any stopping time $T_\textnormal{stop}$ with finite expectation, we  need the following lemma, which is proved in \secref{pflemstopping}. 
\begin{lemma}\label{lem:stopping}
	If $\min\{\tau_j-\lambda_j\} >\epsilon\ $ for some $\epsilon>0$, and $T$ is a random variable supported on $\mathbb{N}$ with finite expectation, then the random variable
	\begin{equation*}
	Y \defn A(T)\equiv \sum_{j\leq T, j\in \nulls} \left( \ident\{P_j \leq \alpha_j \} + \frac{\alpha_j}{(\tau_j-\lambda_j)} \ident\{\lambda_j \leq P_j < \tau_j\} \right)
	\end{equation*}
	also has finite expectation.
\end{lemma}
\noindent

Since $A(T_\textnormal{stop} \land t ) \to A(T_\textnormal{stop})$ almost surely as $t \to \infty$, using \lemref{stopping} and the dominate convergence theorem, we  conclude that \begin{equation}\label{converge}
\EE{A(T_\textnormal{stop} \land t)} \to \EE{A(T_\textnormal{stop})},\ \textnormal{ as }\ t \to \infty.
\end{equation}
Additionally notice that
\begin{equation}\label{inq}
\EE{A(T_\textnormal{stop} \land t )}  = \EE{A(T_\textnormal{stop}\land t) - A(0)} = \EE{(H \cdot A)(t)},   
\end{equation}
where 
\begin{align*}
(H\cdot A)(t)& \defn \sum_{m=1}^{t} H(m)(A(m) - A(m-1)), \textnormal{ and }\ H(t) \defn \ident\{T_\textnormal{stop} \geq t\}.	
\end{align*}
\noindent
Since $T_\textnormal{stop}$ is a stopping tome, it holds that $\{T_\textnormal{stop}\geq t\} = \{T_\textnormal{stop} \leq t\}^{c} \in \Ft$, therefore $H(t+1)$ is measurable with respect to $\mathcal{F}^{t}$. Taking conditional expectation, we  have:
\begin{align*}
&\ \ \ \ \ \EEst{(H \cdot A)(t+1)}{\F^{t}, S_{t+1}} \\ &= \EEst{(H \cdot A)(t)}{\F^{t}, S_{t+1}} + \EEst{H(t+1) (A(t+1)-A(t))}{\F^{t}, S_{t+1}}\\
& \stackrel{(i)}{=}  \EEst{(H \cdot A)(t)}{\F^{t}, S_{t+1}} 
\\& \ \ \ \ + H(t+1) \ident\{t+1 \in \nulls\}\EEst{-\ident\{P_{t+1} \leq \alpha_{t+1} \} + \frac{\alpha_{t+1}}{(\tau_{t+1}-\lambda_{t+1})} \ident\{\lambda_{t+1} < P_{t+1} \leq  \tau_{t+1}\} }{\F^{t}, S_{t+1}}\\
& \stackrel{(ii)}{\geq} \EEst{(H \cdot A)(t)}{\F^{t}, S_{t+1}} + H(t+1)\ident\{t+1 \in \nulls\}(-\alpha_{t+1}/\tau_{t+1} + \alpha_{t+1}/\tau_{t+1}) \ident\{S_{t+1}=1\}\\
& = \EEst{(H \cdot A)(t)}{\F^{t}, S_{t+1}}, 
\end{align*}
where (i) is obtained from the predictability of $H(t+1)$ with respect to $\F^{t}$, and the definition of $A(t)$; and (ii) is obtained using the uniform conservative property  \eqref{condition-conserve-def} of nulls.\\

\noindent
Therefore, additionally applying the law of iterated expectation, we can have that:
\[
\EE{(H \cdot A)(t+1)} \geq \EE{(H \cdot A)(t)}.
\]
Iteratively applying the same argument, we reach the conclusion that, for all $t \in \mathcal{N}:$
\begin{equation}
\EE{(H \cdot A)(t)} \geq 0.
\end{equation}
\noindent
Combining with \eqref{converge} and \eqref{inq}, we  have that, for any stopping time $T_\textnormal{stop}$ with finite expectation, $A(T_\textnormal{stop}) \geq 0$ , which leads to $\mfdr(T_\textnormal{stop}) \leq \alpha$ as we discussed in the beginning.

\subsection{Proof of \lemref{stopping}}\label{sec:pflemstopping}

We prove this lemma using an equivalent form  of Y. Specifically, notice that we can reformulate $Y$ as:
\[
Y = \sum_{j=1}^{\infty}  \left( \ident\{P_j \leq \alpha_j \} + \frac{\alpha_j}{(\tau_j-\lambda_j)} \ident\{\lambda_j < P_j \leq \tau_j\}  \right) \ident\{j\leq T\}.
\]
From the condition that $\min{\{\tau_j-\lambda_j\}} \geq \epsilon$, we  have 
\[\ident\{P_j \leq \alpha_j \} + \frac{\alpha_j}{(\tau_j-\lambda_j)} \ident\{\lambda_j < P_j \leq \tau_j\} \leq 1 + \frac{1}{\epsilon} \defn C\ \textnormal{ for all } j.
\]
Thus, we can bound the expectation of $Y$ as:
\begin{equation*}
\begin{split}
\EE{Y} & = \EE{\sum_{j=1}^{\infty}  \left( \ident\{P_j \leq \alpha_j \} + \frac{\alpha_j}{(\tau_j-\lambda_j)} \ident\{\lambda_j \leq P_j < \tau_j\}  \right) \ident\{j\leq T\}
} \\ & \leq C \sum_{j = 1}^{\infty} \PP{T \geq j} = C\ \EE{T} < \infty.
\end{split}
\end{equation*}
Therefore, we  conclude that $Y$ has finite expectation as claimed.

\section{Proof of \thmref{thmdislord}}\label{sec:pfthmdislord}
Similar to the proof of \thmref{addisthm}, part (a) of \thmref{thmdislord} is proved using the property of uniformly conservative null $p$-values as stated in \eqref{condition-conserve-def}, and the law of iterated expectation. Specifically, conditioning on $\{\Fj, S_j\}$ respectively for each $j \in \nulls$, we have 
\begin{equation*}\label{proof1}
\begin{split}
\EE{|\nulls\cap R(t)|} & =  \sum_{j\leq t, j\in \nulls}\EE{\ident\{P_j \leq \alpha_j\}} = \sum_{j\leq t, j\in \nulls} \EE{\EEst{ \ident\{P_j \leq \alpha_j\}}{S_j, \Fj}}\\
& \stackrel{(i)}{=}  \sum_{j\leq t, j\in \nulls}\EE{\EEst{\ident\{\frac{P_j}{\tau_j} \leq \frac{\alpha_j}{\tau_j}\}}{P_j\leq\tau_j, \Fj}\PPst{S_j = 1}{\Fj}}\\
& \stackrel{(ii)}{\leq} \sum_{j\leq t, j\in \nulls}\EE{\frac{\alpha_j}{\tau_j}\PPst{S_j = 1}{\Fj}}\\
& \stackrel{(iii)}{=}  \sum_{j\leq t, j\in \nulls}\EE{\EEst{ \frac{\alpha_j}{\tau_j}\ident\{P_j\leq \tau_j\}}{P_j\leq\tau_j, \Fj}\PPst{S_j = 1}{\Fj}}\\
& =  \sum_{j\leq t, j\in \nulls}\EE{\EEst{ \frac{\alpha_j}{\tau_j}\ident\{P_j\leq \tau_j\}}{S_j, \Fj}}\\
& \stackrel{(iv)}{=} \sum_{j\leq t, j\in \nulls}\EE{ \frac{\alpha_j}{\tau_j} \ident\{P_j\leq \tau_j\}} = \EE{ \sum_{j\leq t, j\in \nulls}\frac{\alpha_j}{\tau_j} \ident\{P_j\leq \tau_j\}},
\end{split}
\end{equation*}
\noindent
where (i) is true since $\alpha_j < \tau_j$ for any $j$; (ii) is obtained using the uniformly conservative property of null $p$-values; (iii) is true since $\alpha_j$ and $\tau_j$ are both predictable given $\Fj$; and (iv) is obtained using the law of iterated expectation. Therefore, we reach the conclusion that 
\begin{equation}\label{statement}
\EE{|\nulls\cap R(t)|} \leq\EE{\sum_{j \leq t, j\in \nulls}\alpha_j\frac{\ident\{ P_j \leq \tau_j\}}{\tau_j}}.
\end{equation}

Furthermore, since
\[
\fdphat_{\textnormal{\textnormal{D-LORD}}}(t)\leq\alpha \Rightarrow \sum_{j\leq t} \alpha_j \frac{\ident\{P_j\leq\tau_j\}}{\tau_j} \leq \alpha (|R(t)|\vee 1), 
\]
take expectation on each side and use \eqref{statement}, we  obtain $\mfdr(t)\leq \alpha$ with simple rearrangement, which concludes the proof of part (a).\\

\noindent
Additionally, under the independence and monotonicity assumption of part (b), using \lemref{addislem} with simple modification, together with the same trick of taking iterated expectation and repeatedly using the definition of uniformly conservative nulls, we   have the following:
\begin{equation}
\begin{split}
\fdr(t) = \EE{\fdp(t)} &= \EE{\frac{|\nulls\cap R(t)|}{|R(t)\vee 1|}}=\sum_{j\leq t, j\in \nulls}\EE{\frac{\ident\{P_j\leq\alpha_j\}}{|R(t)|\vee 1}}\\
&=  \sum_{j\leq t, j\in \nulls}\EE{\EEst{\frac{\ident\{P_j\leq\alpha_j\}}{|R(t)|\vee 1}}{S_j, \Fj}} \\
&=  \sum_{j\leq t, j\in \nulls}\EE{\EEst{\frac{\ident\{P_j\leq\alpha_j\}}{|R(t)|\vee 1}}{S_j=1, \Fj}\PPst{S_j = 1}{\Fj}} \\
& \leq \sum_{j\leq t, j\in \nulls } \EE{\EEst{\frac{\alpha_j}{\tau_j(|R(t)|\vee 1)}}{S_j = 1, \Fj} \PPst{S_j = 1}{\Fj}}\\
& = \sum_{j\leq t, j\in \nulls } \EE{\EEst{\frac{\alpha_j}{\tau_j(|R(t)|\vee 1)}}{S_j, \Fj}}\\
& = \sum_{j\leq t, j\in \nulls } \EE{\frac{\alpha_j}{\tau_j(|R(t)|\vee 1)}}\\
&\leq \EE{\fdphat_{\textnormal{D-LORD}}(t)}\leq \alpha.
\end{split}
\end{equation}
This concludes the proof of statement (b).

\section{Proof of \thmref{thmdisstorey}}\label{sec:pfthmdisstorey} 
We will prove this theorem using the trick of leave-one-out and the following lemma from \cite{ramdas2019unified}. 
\begin{lemma}\label{lem:lemstorey} (Inverse Binomial Lemma from \cite{ramdas2019unified}) 
	Given a vector $a \defn (a_1, \dots, a_m)\in [0, 1]^m$ , constant $b \in [0, 1]$, and independent Bernoulli variables $Z_i \sim Bernoulli(b)$, the weighted sum $Z = 1 + \sum_{i=1}^{m}a_i Z_i$ satisfies
	\begin{equation}
	\frac{1}{1+b\sum_{i=1}^{m}a_i} \leq \EE{\frac{1}{Z}} \leq  \frac{1}{b(1+\sum_{i=1}^{m}a_i)}.
	\end{equation}
\end{lemma}
\noindent We refer reader to the paper for detailed proof of \lemref{lemstorey}.

\noindent
For a fixed $i \in \nulls \cap \S$, where $\S = \{j : P_j \leq \tau, j \in [n]\}$, we use the leave-one-out trick to define some random variable that is independent with $P_i$,  say $Y^{-i} \defn 1+ \sum_{j\in \nulls, j\neq i} \ident\{ \lambda < P_j \leq  \tau\}$. In this way, for all $j\in \nulls, j\neq i$, $Y^{-i}_j \defn \ident\{ \lambda < P_j \leq  \tau\}$ is stochastically larger than $\text{Bernoulli}(1-\lambda)$ for $j \in \nulls$ conditioning on $P_j \leq \tau$, since the uniformly conservativeness defined in \eqref{conserve-def} implies that 
\[
\PPst{\lambda < P_j \leq  \tau}{P_j \leq \tau}\geq 1-\lambda/\tau. 
\]
Denote $m = |\nulls|$, and $m_S = |\nulls\cap \S|$, let $Z = 1 + \sum_{i=1}^{m_S-1}Z_i$, where $\{Z_i\}_{i=1}^{m_S-1}$ are independent Bernoulli random variables with parameter $1-\lambda/\tau$. Additionally , since $p$-values are independent of each other, we  have
\begin{align*}
\EEst{Y^{-i}}{\S} &= 1 +  \sum_{j\in\nulls \cap \S, j\neq i}\EEst{Y^{-i}_j}{P_j \leq \tau}\\
& \geq 1 + \sum_{j\in\nulls \cap \S, j\neq i}\EE{Z_j} = \EEst{Z}{\S}.	    
\end{align*}

Using \lemref{lemstorey}, we  obtain
\begin{equation}\label{yi}
\EEst{\frac{1}{Y^{-i}}}{\S} \leq \EEst{\frac{1}{Z}}{\S} \leq \frac{1}{(1-\lambda/\tau)|\nulls \cap \S|}.
\end{equation}
Let
\begin{equation}\label{pi}
\widehat{\pi}_0^{-i} \defn \frac{1+\sum_{j \leq n, j\neq i} \ident\{\lambda<P_j\leq \tau\}}{n(\tau-\lambda)}.
\end{equation}
It is easy to see that $ \widehat{\pi}_0^{-i} \geq  \frac{Y^{-i}}{n(\tau-\lambda)}$. Together with \eqref{yi} and \eqref{pi}, we  obtain
\begin{equation}\label{pi}
\EEst{\frac{1}{\widehat{\pi}_0^{-i}}}{\S} \leq n(\tau-\lambda)\EEst{\frac{1}{Y^{-i}}}{\S} \leq \frac{n\tau}{|\nulls \cap \S|}.
\end{equation}
Using the definition of $\fdphat_\textnormal{D-StBH}$ in \eqref{disstbh}, and the uniform conservativeness of $p$-values, we have the following:
\begin{equation*}
\begin{split}
\EEst{\frac{\sum_{i\in \nulls \cap \S}\ident\{P_i\leq \widehat{s}\}}{(\sum_{i}\ident\{P_i\leq \widehat{s}\})\vee 1}}{\S} & 
\stackrel{(i)}{\leq} \EEst{\frac{\alpha\sum_{i\in \nulls \cap \S}\ident\{P_i\leq \widehat{s}\}}{n \widehat{\pi}_0 \widehat{s}}}{\S}\\
&= \frac{\alpha}{n} \EEst{\sum_{i\in\nulls \cap \S}\frac{\ident\{P_i \leq \widehat{s}\}}{\widehat{\pi}_0 \widehat{s}}}{\S}
\stackrel{(ii)}{=} \frac{\alpha}{n} \EEst{\sum_{i\in\nulls \cap \S}\frac{\ident\{P_i \leq \widehat{s}\}}{\widehat{\pi}_0^{-i} \widehat{s}}}{\S}\\
&	\stackrel{(iii)}{=}\frac{\alpha}{n} \EEst{\sum_{i\in\nulls \cap \S}\frac{1}{\widehat{\pi}_0^{-i}}\EEst{\frac{\ident\{P_i \leq \widehat{s}\}}{\widehat{s}}}{\mathcal{P}^{-i}, \S}}{\S}\\
&	\stackrel{(iv)}{=}\frac{\alpha}{n}\EEst{\sum_{i\in\nulls \cap \S} \frac{1}{\widehat{\pi}_0^{-i}}\EEst{\frac{\ident\{P_i \leq \widehat{s}\}}{\widehat{s}}}{\mathcal{P}^{-i}, P_i \leq \tau}}{\S}\\
&	\stackrel{(v)}{\leq} \frac{\alpha}{n}\EEst{\sum_{i\in\nulls \cap \S}\frac{1}{\tau\widehat{\pi}_0^{-i}}}{\S} = \frac{\alpha}{n}\sum_{i\in\nulls \cap \S}\EEst{\frac{1}{\tau\widehat{\pi}_0^{-i}}}{\S} \stackrel{(vi)}{\leq}\alpha, 
\end{split}
\end{equation*}
where (i) follows from the condition $\fdphat_\textnormal{D-StBH} \leq \alpha$; (ii) is true since $\widehat{\pi}_0^{-i} = \widehat{\pi}_0$ given $\ident\{P_i\leq \widehat{s}\} = 1$, using the fact that $\widehat{s} \leq \lambda$; (iii) is true since conditioning on $\mathcal{P}_i$ fully determines $\widehat{\pi}_0^{-i} $; (iv) follows from the fact that $\widehat{s} \leq \lambda$; and (v) is obtained by noticing  $\widehat{s}$ is coordinatewise nondecreasing in $P_i$ for each $i$, and using the lemma 1 in \cite{ramdas2019unified}; and the final step (vi) follows from \eqref{pi}. Therefore, we obtain that $\EEst{\fdp}{\S}\leq \alpha$.  
Taking expectation with regard $\S$ on both side, we  have $\fdr \leq \alpha$ as claimed.

\section{Proof of \thmref{asyncaddis}}\label{sec:pfasync}
\thmref{asyncaddis} is proved using similar technique in the proof of \thmref{addisthm}, we present the proof here for completeness. Similarly, we need the following lemma for the proof, which is proved in \secref{pfasynclem}.

\begin{lemma}\label{lem:asynclem}
	Assume that the $p$-values $P_1,P_2,\dots$ are independent and let $g: \{0,1\}^T \to \mathbb{R}$ be any coordinatewise nondecreasing function. Further, assume that $\alpha_t$,$\lambda_t$ and $1-\tau_t$ are all monotonic functions of the past as defined in \secref{async}, while satisfying the constraints $\alpha_t \leq \lambda_t < \tau_t$ for all $t$. Then, for any index $t \leq T$ such that $H_t \in \nulls$, we have:
	\begin{align*}
	\EEst{\frac{\alpha_t\ident\{\lambda_t<P_t\leq\tau_t\}}{(\tau_t-\lambda_t)(g(|\R|_{1:T})\vee 1)}}{\F^{E_t -1}, S_t = 1} & \geq \EEst{\frac{\alpha_t }{\tau_t (g(|\R|_{1:T})\vee 1)}}{\F^{E_t -1}, S_t = 1}
	\\&\geq\EEst{\frac{\ident\{P_t\leq \alpha_t \}}{g(|\R|_{1:T})\vee 1}}{\F^{E_t -1}, S_t = 1},
	\end{align*}
	where $|\R|_{1:T}\defn \{|\R_1|, \dots, |\R_T|\}$.
\end{lemma}

\noindent Denote $\R(t) = \{i: P_i \leq \alpha_i, E_i \leq t\}$,  we have the following:
\begin{equation*}
\begin{split}
\EE{|\nulls\cap \R(t)|}& =  \EE{\sum_{E_j\leq t, j\in \nulls}\ident\{P_j \leq \alpha_j\}} \stackrel{(i)}{\leq}  \sum_{j\leq t, j\in \nulls}\EE{\ident\{P_j \leq \alpha_j\}}\\
& \stackrel{(ii)}{=}  \sum_{j\leq t, j\in \nulls}\EE{\EEst{\ident\{P_j \leq \alpha_j\}}{\F^{E_j -1}, S_j}}\\
& \stackrel{(iii)}{=}  \sum_{j\leq t, j\in \nulls}\EE{\EEst{\ident\{\frac{P_j}{\tau_j} \leq \frac{\alpha_j}{\tau_j}\}}{P_j\leq\tau_j, \F^{E_j -1}} \PPst{S_j = 1}{\F^{E_j -1}}}\\
& \stackrel{(iv)}{\leq} \sum_{j\leq t, j\in \nulls}\EE{ \frac{\alpha_j}{\tau_j}\PPst{S_j = 1}{\F^{E_j -1}}},
\end{split}
\end{equation*}
where step (i) is true since the set of rejections by time $t$ could be at most $[t]$; and (ii) is obtained via taking iterated expectation by conditioning on $\{\F^{E_j-1}, S_j\}$ respectively for each $j \in \nulls$;  and (iii) is true since $\alpha_j \leq \tau_j$; and finally, step (iv) follows from the uniformly conservativeness of nulls. Next, notice that


\begin{equation*}
\begin{split}
&\ \ \ \ \ \ \sum_{j\leq t, j\in \nulls}  \EE{\frac{\alpha_j}{\tau_j}\PPst{S_j =1}{\F^{E_j -1}}}\\
& \stackrel{(v)}{\leq} \sum_{j\leq t, j\in \nulls}\EE{\frac{\alpha_j}{\tau_j}\EEst{\frac{\ident\{\lambda_j<P_j\}}{(1-\lambda_j/\tau_j)}}{P_j \leq \tau_j, \F^{E_j -1}}\PPst{S_j =1}{\F^{E_j -1}}}\\
& = \sum_{j\leq t, j\in \nulls}\EE{\frac{\alpha_j}{\tau_j}\EEst{\frac{\ident\{\lambda_j<P_j \leq \tau_j\}}{(1-\lambda_j/\tau_j)}}{S_j, \F^{E_j -1}}}\\
& = \sum_{j\leq t, j\in \nulls}\EE{\alpha_j\frac{\ident\{\lambda_j<P_j\leq \tau_j\}}{(\tau_j-\lambda_j)}}
\end{split}
\end{equation*}
where (v) is true because of the uniformly conservativaness of null $p$-values, and the last two equalities use the predictability of $\alpha_j$ and $\tau_j$ with regard $\F^{E_j -1}$. Then, by removing some constrains on the index, and applying the condition that $\fdphat_{\textnormal{ADDIS}_{\textnormal{async}}} \leq \alpha$, one  obatin
\begin{equation*}
\begin{split}
&\sum_{j\leq t, j\in \nulls}\EE{\alpha_j\frac{\ident\{\lambda_j<P_j\leq \tau_j\}}{(\tau_j-\lambda_j)}}\\
\leq &\ \sum_{j\leq t}\ \EE{\frac{\alpha_j}{(\tau_j-\lambda_j)}(\ident\{\lambda_j<P_j \leq \tau_j, E_j < t\} + \ident \{E_j \geq t\})}\\
\leq &\ \alpha \ \EE{\left(\sum_{j\leq t} \ident\{P_j \leq \alpha_j, E_j < t\}\right)\vee 1} = \ \alpha\ \EE{|\R(t)|\vee 1},
\end{split}
\end{equation*}
Therefore, we have
\[
\EE{|\nulls \cap \R(t)|} \leq \alpha \EE{|\R(t)|\vee 1}.
\]
After rearranging the terms above, we have $\mfdr(t) \leq \alpha$, as claimed. Therefore, we finished the proof of first part of \thmref{asyncaddis}.\\

\noindent
Using the same tricks of taking iterated expectation, we  have the following:
\begin{equation}\label{asynceq1}
\begin{split}
\fdr(t) = \EE{\fdp(t)} 
&= \EE{\frac{|\nulls\cap\R(t)|}{|\R(t)|\vee 1}} =\EE{ \frac{\sum_{E_j \leq t, j\in \nulls} \ident\{P_j\leq\alpha_j\}}{|\R(t)|\vee 1}}\\
&\leq \EE{ \frac{\sum_{j \leq t, j\in \nulls} \ident\{P_j\leq\alpha_j\}}{|\R(t)|\vee 1}} = \sum_{j \leq t, j\in \nulls} \EE{ \frac{ \ident\{P_j\leq\alpha_j\}}{|\R(t)|\vee 1}}\\
&=  \sum_{j\leq t, j\in \nulls}\EE{\EEst{\frac{\ident\{P_j\leq\alpha_j\}}{|\R(t)|\vee 1}}{S_j, \F^{E_j -1}}}. \\
\end{split}
\end{equation}
Under additional assumptions about the  independence of $p$-values and monotonicity of $\alpha_t$ and $\lambda_t$ for each $t \in N$, and notice that $|\R(t)| = \sum_{i \leq t, E_i < t} R_i = \sum_{i < t} |\R_i|$ is coordinatewise nondecreasing function of $|\R|_{1:t}$, we  apply \lemref{asynclem} to the RHS of \eqref{asynceq1} to obtain the following: 

\begin{equation}\label{asynceq2}
\begin{split}
& \sum_{j\leq t, j\in \nulls}\EE{\EEst{\frac{\ident\{P_j\leq\alpha_j\}}{|\R(t)|\vee 1}}{S_j, \F^{E_j -1}}}\\
= & \sum_{j\leq t, j\in \nulls}\EE{\EEst{\frac{\ident\{P_j\leq\alpha_j\}}{|\R(t)|\vee 1}}{S_j = 1, \F^{E_j -1}}\PPst{S_j = 1}{\F^{E_j -1}}} \\
\leq & \sum_{j\leq t, j\in \nulls}\EE{\EEst{\frac{\alpha_j}{\tau_j(|\R(t)|\vee 1)}}{S_j = 1, \F^{E_j -1}}\PPst{S_j = 1}{\F^{E_j -1}}}\\
\leq & \sum_{j\leq t, j\in \nulls}\EE{ \EEst{\frac{\alpha_j}{\tau_j(|\R(t)|\vee 1)}\frac{\ident\{\lambda_j<P_j\leq\tau_j\}}{1-\lambda_j/\tau_j}}{S_j = 1, \F^{E_j -1}}\PPst{S_j = 1}{\F^{E_j -1}}}\\
\end{split}
\end{equation}
Once again using the fact that $\alpha_j \leq \tau_j$ for all $j$, and the law of iterated expectation, the RHS of \eqref{asynceq2} equals 
\begin{equation}\label{asynceq3}
\begin{split}	
& \sum_{j\leq t, j\in \nulls}\EE{ \EEst{\frac{\alpha_j}{\tau_j(|\R(t)|\vee 1)}\frac{\ident\{\lambda_j<P_j\leq\tau_j\}}{1-\lambda_j/\tau_j}}{S_j, \F^{E_j -1}}}\\
= & \sum_{j\leq t, j\in \nulls}\EE{\frac{\alpha_j}{\tau_j(|\R(t)|\vee 1)} \frac{\ident\{\lambda_j<P_j\leq\tau_j\}}{(1-\lambda_j/\tau_j)}}\\
\leq & \sum_{j\leq t}\EE{\frac{1}{|\R(t)|\vee 1} \frac{\alpha_j}{(\tau_j-\lambda_j)}(\ident\{\lambda_j<P_j\leq\tau_j , E_j < t\} + \ident\{E_j \geq t\})}\\
= &\EE{\fdphat_{\textnormal{ADDIS}_\textnormal{async}}(t)}\leq \alpha.
\end{split}
\end{equation}
Therefore, combining \eqref{asynceq1}, \eqref{asynceq3}, and \eqref{asynceq3}, we  conclude $\fdr(t) \leq \alpha$. This finishes the proof of the second part of the theorem.

\subsection{Proof of  \lemref{asynclem}}\label{sec:pfasynclem}
Similar to the proof of \lemref{addislem}, we prove this lemma by constructing a hallucinated vector. Specifically, to prove the first part of the inequality, for any fixed $t\in \N$, for all $i \in \N$, let $\widetilde{P}_i = \tau_i \cdot \ident\{i=t\} + P_i \cdot \ident\{i\neq t\}$, and keep the finish times for all the tests unchanged. Then we denote the testing levels, candidate levels and selected levels  resulted from the hallucinated $\{\widetilde{P}_i\}$ as $\{\widetilde{\alpha}_i\}$, $\{\widetilde{\lambda}_i\}$ and $\{\widetilde{\tau}_i\}$ respectively. Correspondingly, we let
\[
\widetilde{S}_i = \ident\{\widetilde{P}_i\leq \widetilde{\tau}_i\},\ \ \  \widetilde{C}_i=\ident\{\widetilde{P}_i\leq\widetilde{\lambda}_i\},\ \ \ \ \widetilde{R}_i = \ident\{\widetilde{P}_i\leq \widetilde{\alpha}_i\}.
\]

\noindent
Given $\lambda_t <P_t \leq \tau_t$, we have $\widetilde{S}_t = S_t = 1, \widetilde{R}_t = R_t = 0, \widetilde{C}_t = C_t = 0$. This implies $\R_{1:T} = \widetilde{\R}_{1:T}$. We then obtain the following:
\begin{equation*}
\begin{split}
&\ \ \ \ \ \EEst{\frac{\alpha_t\ident\{\lambda_t<P_t\leq \tau_t\}}{(\tau_t-\lambda_t)(g(|\R|_{1:T})\vee 1)}}{S_t = 1, \F^{E_t -1}} = \EEst{\frac{\alpha_t\ident\{\lambda_t<P_t\leq \tau_t\}}{(\tau_t-\lambda_t)(g(\widetilde{|\R|}_{1:T})\vee 1)}}{S_t = 1, \F^{E_t -1}}\\
& \stackrel{(i)}{=} \EEst{\frac{\alpha_t}{\tau_t (g(\widetilde{|\R|}_{1:T})\vee 1)}}{S_t = 1, \F^{E_t -1}}\EEst{\frac{\ident\{\lambda_t<P_t\leq \tau_t\}}{(1-\lambda_t/\tau_t)(g(\widetilde{|\R|}_{1:T})\vee 1)}}{S_t = 1, \F^{E_t -1}}\\
&\stackrel{(ii)}{\geq} \EEst{\frac{\alpha_t}{\tau_t (g(\widetilde{|\R|}_{1:T})\vee 1)}}{S_t = 1, \F^{E_t -1}}\stackrel{(iii)}{\geq} \EEst{\frac{\alpha_t}{\tau_t (g(|\R|_{1:T})\vee 1)}}{S_t = 1, \F^{E_t -1}},
\end{split}
\end{equation*}
where (i) is obtained from the fact that $\widetilde{\R}_{1:T}$ is independent of $P_t$, and (ii) is true because of the uniformly conservativaness of null $p$-values, and (iii) is true since $\widetilde{R}_i \subseteq R_i$ for all $i$ given $S_t = 1$ using the similar logic in the proof of \lemref{addislem} in \secref{pfaddislem}, that is utilizing the monotonicity assumptions of $\alpha_t$, $\lambda_t$, and $\tau_t$.\\

\noindent
Similarly, for the second part of the inequality, we construct $\widetilde{P}_i = P_i \cdot \ident\{i=t\}$, and keep the finish times for all the tests unchanged, while we define $\widetilde{\alpha}_t$, $\widetilde{\lambda}_t$, $\widetilde{\tau}_t$ and $\widetilde{R}_t$, $\widetilde{C}_t$, $\widetilde{S}_t$ in the same way as in the proof of the first part. Notice that $\widetilde{\R}_{1:T}$ is independent of $P_t$, and that given $P_t\leq \alpha_t$, we have $\widetilde{S}_t = S_t = 1, \widetilde{R}_t=\widetilde{C}_t=R_t= C_t = 1$, which leads to $\R_{1:T} = \widetilde{\R}_{1:T}$. Then we have the following:
\begin{equation*}
\begin{split}
&\EEst{\frac{\ident\{P_t\leq \alpha_t\}}{g(|\R|_{1:T})\vee 1}}{S_t = 1, \F^{E_t -1}} = \EEst{\frac{\ident\{P_t\leq \alpha_t\}}{g(\widetilde{|\R|}_{1:T})\vee 1}}{S_t = 1, \F^{E_t -1}}\\
&\leq\EEst{\frac{\alpha_t}{\tau_t (g(\widetilde{|\R|}_{1:T})\vee 1)}}{S_t =1, \F^{E_t -1}} \leq\EEst{\frac{\alpha_t}{\tau_t (g(|\R|_{1:T})\vee 1)}}{S_t = 1, \F^{E_t -1}}.
\end{split}
\end{equation*}
This concludes the whole proof \lemref{asynclem}.

\section{An equivalent form of ADDIS$^{*}$ algorithm}\label{sec:addisalgoeq}
\begin{algorithm}[H]
	\KwIn{FDR level $\alpha$, discarding threshold $\tau \in (0,1]$, candidate threshold $\lambda \in [\alpha,\tau)$, sequence $\{\gamma_j \}_{j=0}^{\infty}$ which is nonnegative, nonincreasing and sums to one, initial wealth $W_0 \leq \alpha$.}
	\For{$t=1, 2, \dots$}{
		\If{$P_t > \tau$}{Discard $P_t$ and move to next round.}
		\Else{
			Reject the $t$-th null hypothesis if  $P_t/\tau \leq \alpha_t$, where\\
			${\alpha}_{t}  \defn (1-\lambda)\left(W_0\gamma_{S^t - C_{0+}} +(\alpha-W_0)\gamma_{S^t-\kappa_1^{*}-C_{1+}}  + \alpha\sum_{j\geq 2} \gamma_{S^t - \kappa_j^{*}-C_{j+}}\right).$\\
			Here, $\ \ S^{t} = \sum_{i< t} \ident\{ P_i \leq \tau\} $, $ \ \ \ \ C_{j+} = \sum_{i=\kappa_j+1}^{t-1}\ident\{ P_i \leq \lambda\}$, \\
			$ \ \ \ \  \ \ \ \ \ \ \ \ \kappa_j = \min \{i\in [t-1]: \sum_{k\leq i} \ident\{ P_k \leq \alpha_k\} \geq j\}$, $ \ \ \ \  \kappa_j^{*} = \sum_{i \leq \kappa_j} \ident\{ P_i \leq \tau\} $.}
	}
	\caption{The $\textnormal{ADDIS}^{*}$ algorithm with explicit use of discarding}\label{addisalgoeq}
\end{algorithm}

\section{Heatmap of $g \circ F$}\label{sec:secgf}
\begin{figure}[H]
	\centering
	\begin{subfigure}{0.33\textwidth}
		\centering
		\includegraphics[width=\linewidth]{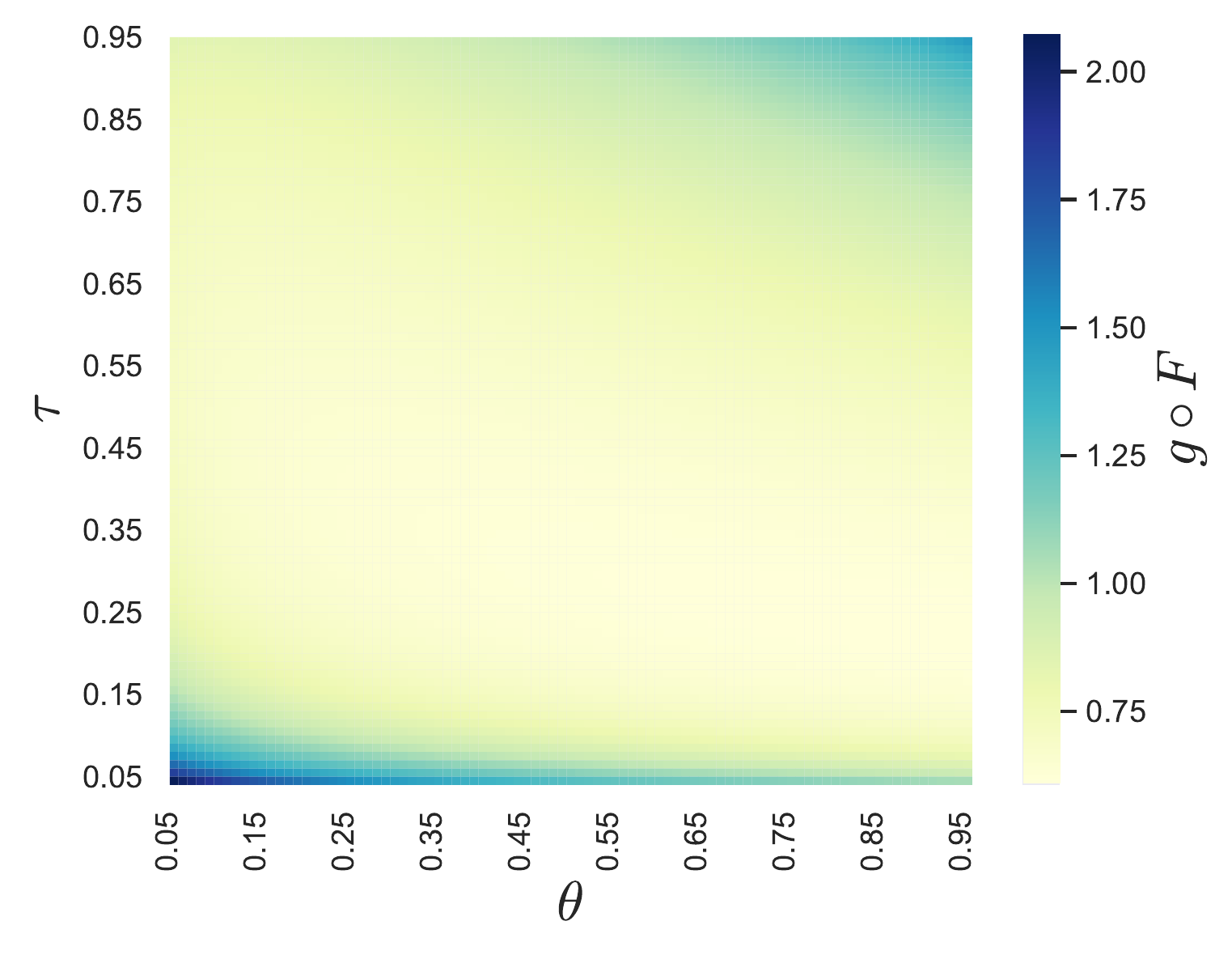}
		\caption{}
	\end{subfigure}%
	\begin{subfigure}{0.33\textwidth}
		\centering
		\includegraphics[width=\linewidth]{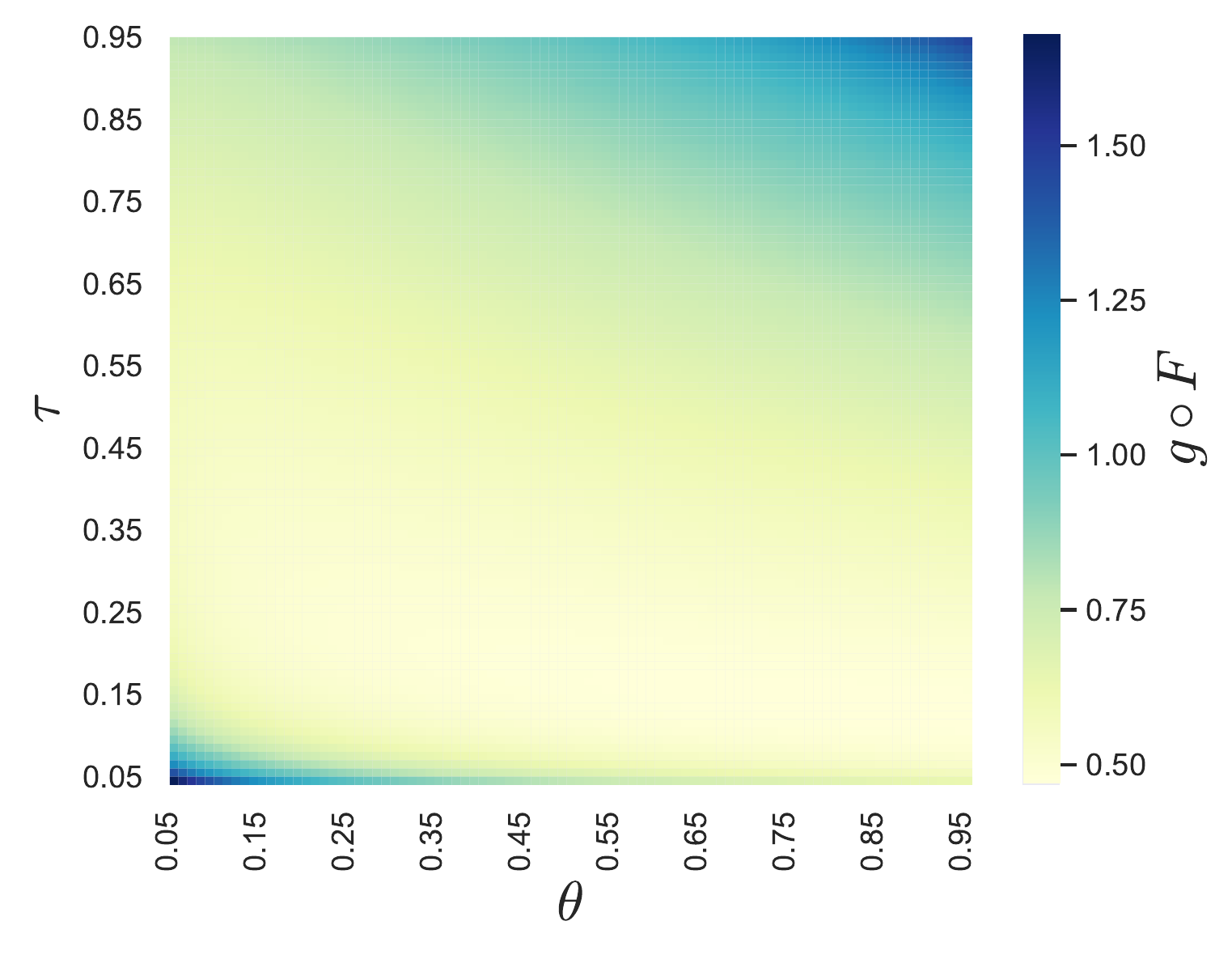}
		\caption{}
	\end{subfigure}	%
	\begin{subfigure}{0.33\textwidth}
		\centering
		\includegraphics[width=\linewidth]{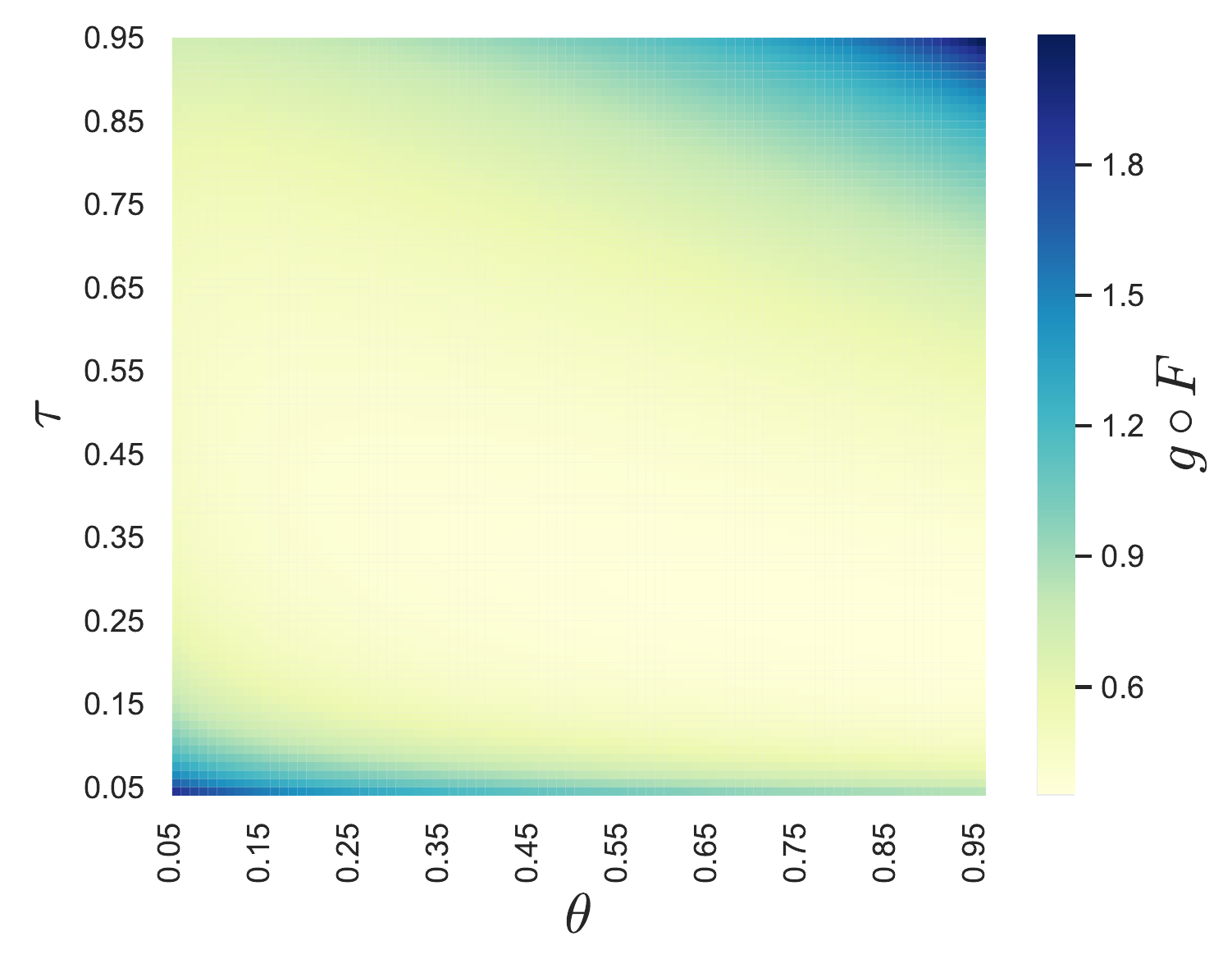}
		\caption{}
	\end{subfigure}\\
	\begin{subfigure}{0.33\textwidth}
		\centering
		\includegraphics[width=\linewidth]{newideal_tradeoff_MN-1.0_MA3.0_pi0.2.pdf}
		\caption{}
	\end{subfigure}%
	\begin{subfigure}{0.33\textwidth}
		\centering
		\includegraphics[width=\linewidth]{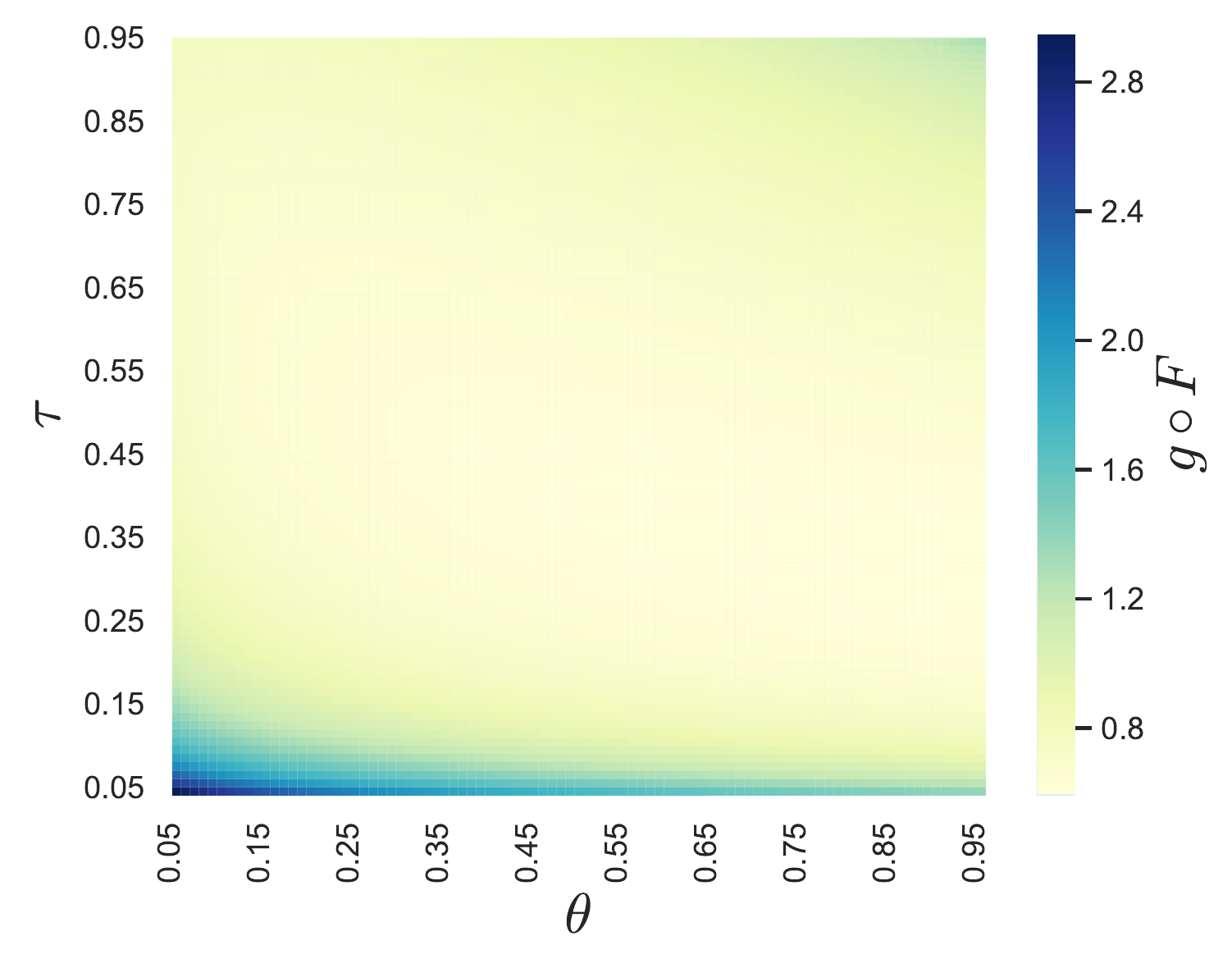}
		\caption{}
	\end{subfigure}%
	\begin{subfigure}{0.33\textwidth}
		\centering
		\includegraphics[width=\linewidth]{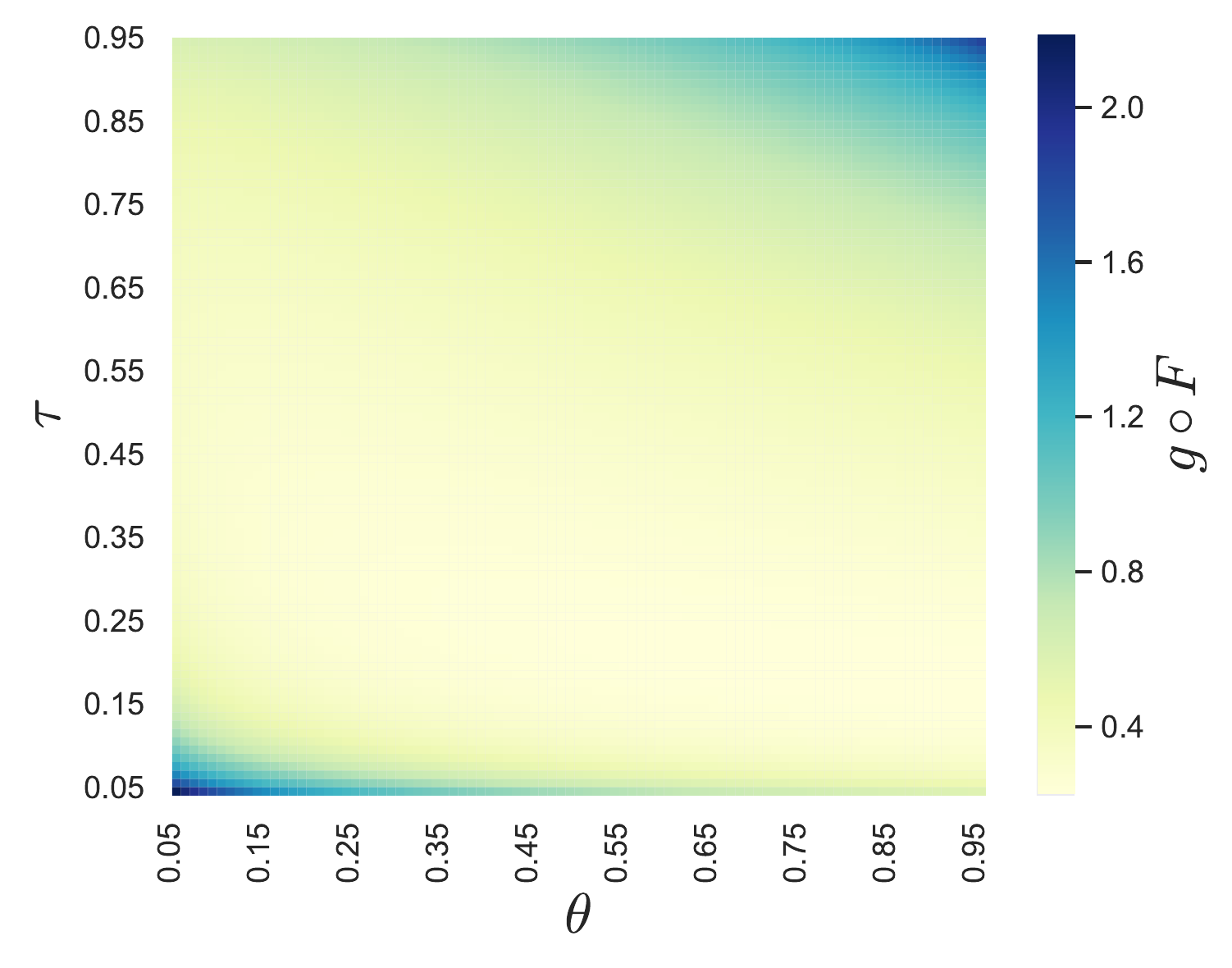}
		\caption{}
	\end{subfigure}\\
	\begin{subfigure}{0.33\textwidth}
		\centering
		\includegraphics[width=\linewidth]{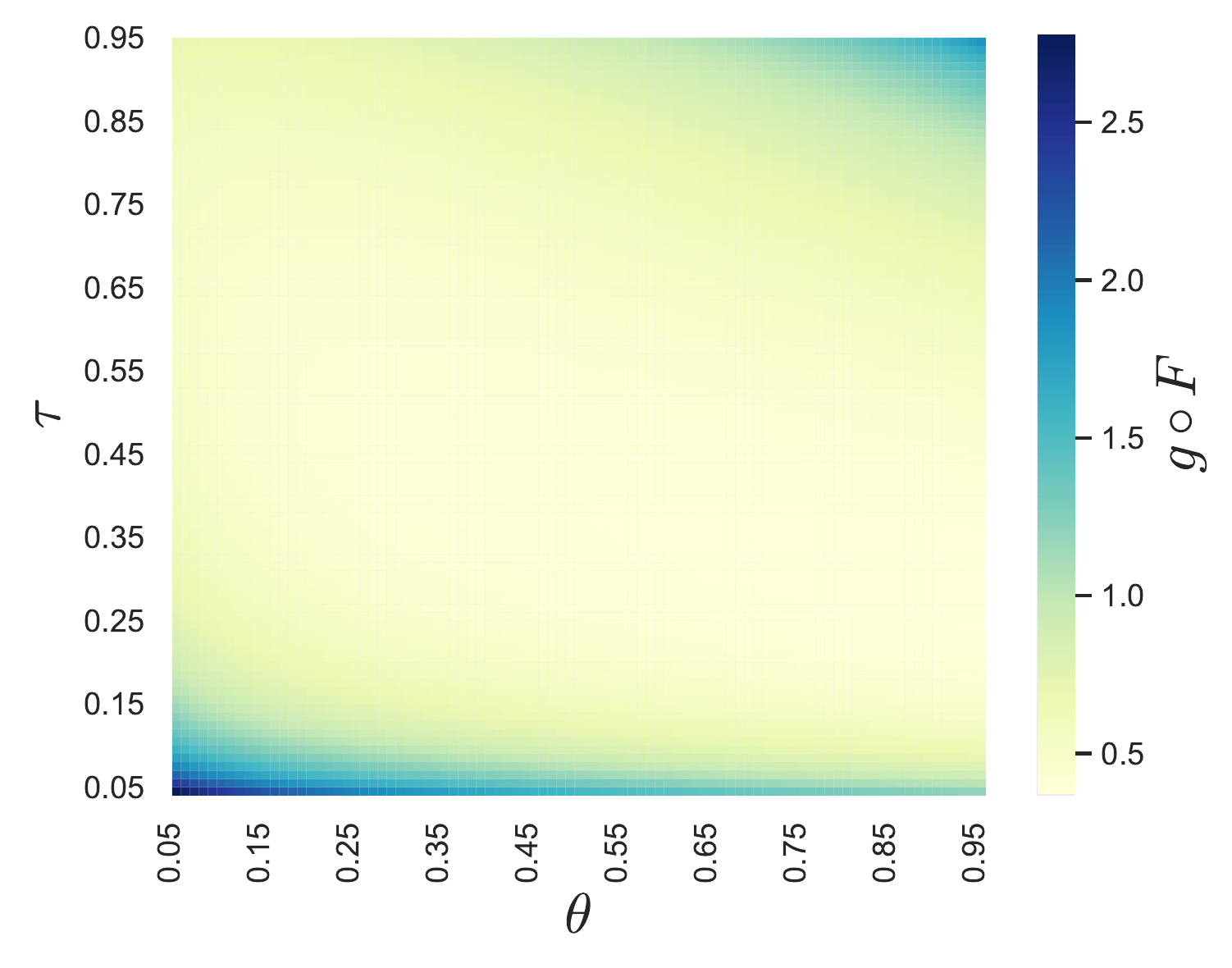}
		\caption{}
	\end{subfigure}%
	\begin{subfigure}{0.33\textwidth}
		\centering
		\includegraphics[width=\linewidth]{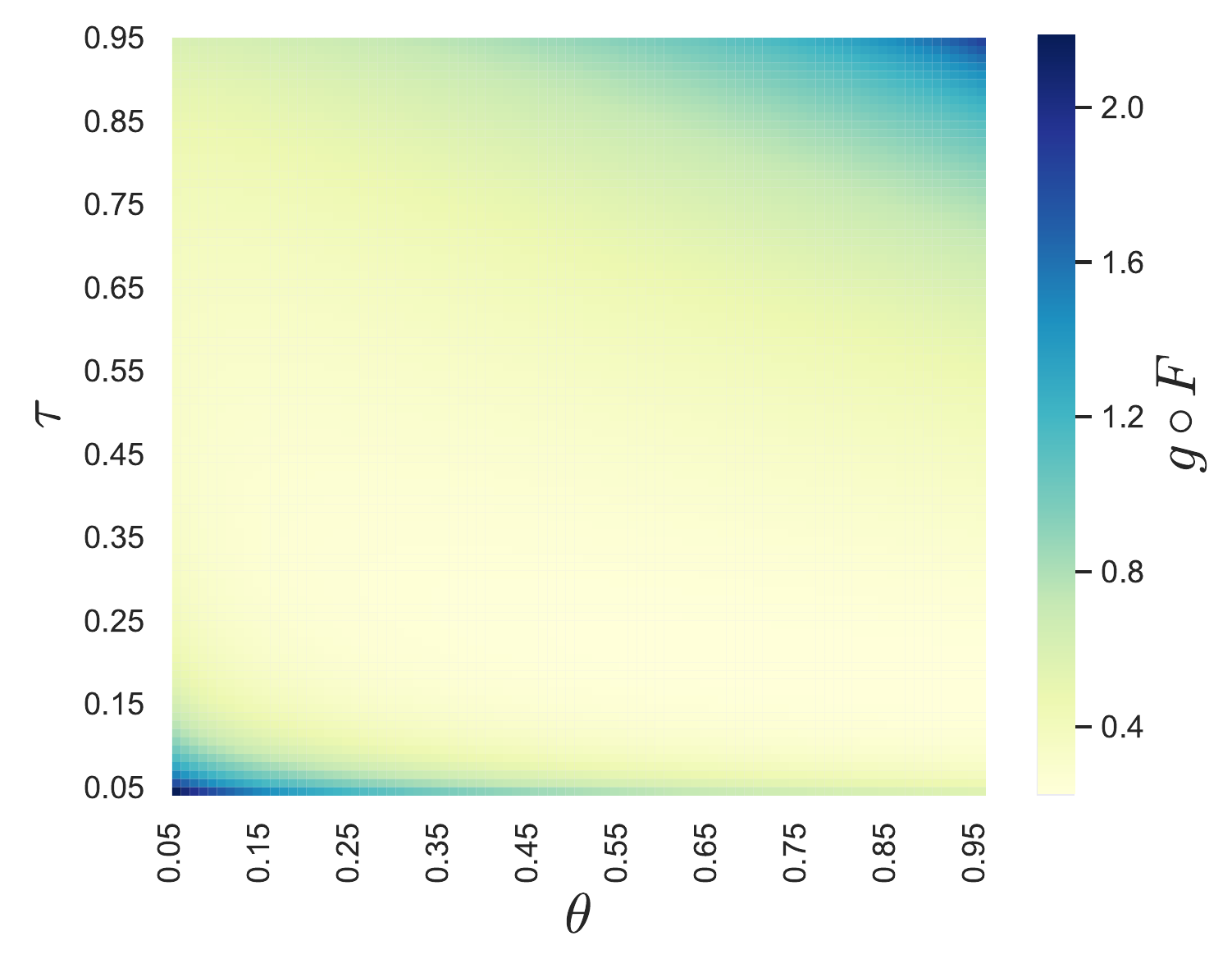}
		\caption{}
	\end{subfigure}
	\caption{The heatmap of function $g\circ F$, where $F$ is the CDF of $p$-values drawn as described in \secref{simu} with $\mu_N = -0.5, \mu_A = 2, \pi_A = 0.2$ for plot (a); $\mu_N = -0.5, \mu_A = 3, \pi_A = 0.2$ for plot (b); $\mu_N = -1, \mu_A = 2, \pi_A = 0.2$ for plot (c); $\mu_N = -1, \mu_A = 3, \pi_A = 0.2$ for plot (d); $\mu_N = -0.5, \mu_A = 2, \pi_A = 0.3$ for plot (e); $\mu_N = -0.5, \mu_A = 3, \pi_A = 0.3$ for plot (f); $\mu_N = -1, \mu_A = 2, \pi_A = 0.3$ for plot (g); $\mu_N = -1, \mu_A = 3, \pi_A = 0.3$ for plot (h).}\label{fig:gf}
\end{figure}

Here we show the heatmap of $g \circ F$ versus $\theta\defn \lambda/\tau$ and $\tau$ given different choices of $F$. Specifically, we let $F$ be the CDF of all $p$-values (nulls and alternatives taken together) drawn as described in \secref{simu}, with different choices of $\mu_N, \mu_A$, and $\pi_A$. In \figref{gf}, we show results for $\mu_N \in \{-0.5, -1\}$,  $\mu_A \in \{2, 3\}$, and $\pi_A \in \{0.2, 0.3\}$ respectively, which are some reasonably common settings that one may expect in practice. We  see that the heatmap of $g \circ F$ demonstrates the same consistent pattern across different choices of $F$.

\end{appendices}

\end{document}